\documentstyle[epsfig,plenum,12pt]{book}
\begin{document}

\newcommand{\be}{\begin{eqnarray}}
\newcommand{\ee}{\end{eqnarray}}

\chapter{UNIVERSAL FLUCTUATIONS IN DIRAC SPECTRA}

\author{Jacobus Verbaarschot}

\affiliation{Department of Physics and Astronomy\\
University at Stony Brook, SUNY\\
Stony Brook, NY 11794, USA}
 
\section{INTRODUCTION}

It is generally accepted that QCD with two massless quarks undergoes a chiral
phase transition (see reviews by
DeTar\refnote{\cite{DeTar}}, Ukawa\refnote{\cite{Ukawa}} and 
Smilga\refnote{\cite{Smilref}} for recent results on this topic). 
This leads to important observable signatures in
the real world with two light quarks.
The order parameter of the chiral phase transition is the chiral condensate
$\langle \bar \psi \psi \rangle$. 
Below the critical temperature chiral symmetry is spontaneously broken with
$\langle \bar \psi \psi \rangle \ne 0$. One consequence is that parity doublets
are absent in the hadronic spectrum. For example, the pion mass
and the $\delta$ mass are very different. To a good approximation,
as dictated by the Goldstone theorem, the pion is massless.
According to lattice QCD simulations, chiral symmetry
is restored for $T> T_c$ where $\langle \bar \psi \psi \rangle= 0$. 
For two light flavors the critical temperature is expected to 
be $T_c \approx 140\, MeV$.
In the restored phase, parity doublets are present, and 
a massive  pion is degenerate with the $\sigma$-meson.
It is well known that the QCD Lagrangian has two chiral symmetries:
the $U_A(1)$ symmetry and the $SU(N_f)\times SU(N_f)$ symmetry. As was
in particular pointed out by Shuryak\refnote{\cite{Edwardaxial}} not
necessarily both symmetries are restored at the same 
temperature\refnote{\cite{DeTarU1}}. This may  lead
to interesting physical consequences. 

According to the Banks-Casher formula\refnote{\cite{Banks-Casher}}, the
chiral condensate
is directly related to the average spectral density  
of the Dirac operator near zero virtuality.  
However, the eigenvalues of the Dirac operator fluctuate about their average
position over the ensemble of gauge field configurations. 
The main question we wish to address in these lectures is to what
extent such fluctuations are universal. If that is the case, they 
do not depend on the full QCD dynamics and can
be obtained from a much simpler chiral Random Matrix Theory (chRMT) 
with the global symmetries of the QCD partition function. 

This conjecture has its origin in the study of 
spectra of complex systems\refnote{\cite{bohigas}}. According to the Bohigas
conjecture, spectral correlations of classically chaotic quantum systems
are given by RMT. A first argument in favor of 
universality in Dirac spectra
came from the analysis of the finite volume QCD partition 
function\refnote{\cite{LS}}.
As has been shown by Gasser and Leutwyler\refnote{\cite{GL}},
for box size $L$ in the range 
\be
1/\Lambda \ll L \ll 1/m_\pi,
\label{range}
\ee
($\Lambda$ is a typical hadronic scale and $m_\pi$ is the pion mass)
the mass dependence of the QCD partition function is completely
determined by its global symmetries. As a consequence, fluctuations of 
Dirac eigenvalues near zero virtuality are constrained by,
but not determined by, an infinite family of sum 
rules\refnote{\cite{LS}} (also called Leutwyler-Smilga sum rules). 
For example, the simplest Leutwyler-Smilga sum rule can be obtained from the
microscopic spectral density\refnote{\cite{SVR}} 
(the spectral density near zero virtuality on 
a scale of a typical eigenvalue
spacing). On the other hand, the infinite family of Leutwyler-Smilga
sum rules is not sufficient to determine the microscopic spectral density.
The additional ingredient is
universality. A priori there is no reason that fluctuations of 
Dirac eigenvalues are in the same universality class as chRMT. Whether or not
QCD is inside this class is a dynamical question that can only be answered
by full scale lattice QCD simulations. However, the confidence in an 
affirmative answer to this question is greatly enhanced by 
universality studies within chiral Random Matrix Theory. 
The aim of such studies is to show that
spectral fluctuations do not depend of the details of the probability 
distribution. Recently, it has been shown that the microscopic spectral
density is universal for a wide class of probability 
distributions\refnote{\cite{Nishigaki-uni,
Damgaard,brezin-hikami-zee,GWu,Sener1,Seneru}}. 
We will give an extensive review of these important new results.

The fluctuations of Dirac eigenvalues near zero virtuality
are directly related to the approach to the thermodynamic limit of
the chiral condensate. In particular, knowledge of the microscopic 
spectral density provides us with a quantitative 
explanation\refnote{\cite{VPLB}} of
finite size corrections to the valence quark mass of 
dependence of the chiral condensate\refnote{\cite{Christ}}. 

Because of the $U_A(1)$ symmetry of the Dirac operator two types
of spectral fluctuations can be distinguished. Spectral fluctuations 
near zero virtuality and spectral fluctuations in the bulk of the spectrum
(Actually, there is a third type: spectral fluctuations near the end-points
of lattice QCD Dirac spectra. However, this region of the spectrum is
completely unphysical, and it will not be considered in these lectures.)

Recently, it has become possible to obtain $all$ eigenvalues of the lattice
QCD Dirac operator on reasonably large 
lattices\refnote{\cite{Kalkreuter,berbenni}}, making 
a direct verification of the above conjecture possible. 
This is one of the main objectives of these lectures. This is easiest
for correlations in the bulk of the spectrum. Under the assumption
of spectral ergodicity\refnote{\cite{pandey}}, 
eigenvalue correlations can be studied by spectral 
averaging instead of ensemble averaging\refnote{\cite{HV,HKV}}. 
On the other hand, in order to 
study the microscopic spectral density, a very large number of independent
gauge field configurations is required. First lattice results confirming
the universality of the microscopic spectral density have been obtained
recently\refnote{\cite{berbenni}}.

At this point I wish to stress that there are two different types of 
applications
of Random Matrix Theory. In the first type, fluctuations
of an observable are related to its average. Because of universality
it is possible to obtain exact results. In general, the average of 
an observable is not given by Random Matrix Theory. There are many examples
of this type of universal fluctuations ranging
from atomic physics to quantum field theory (a recent comprehensive review
was written by Guhr, M\"uller-Groeling and
Weidenm\"uller\refnote{\cite{hdgang}}). Most of the examples are related to
fluctuations of eigenvalues. Typical examples are
nuclear spectra\refnote{\cite{Haq}},
acoustic spectra\refnote{\cite{Guhr}}, resonances in resonance 
cavities\refnote{\cite{Koch}},
$S$-matrix fluctuations\refnote{\cite{ERICSON,WEIDI}} and
universal conductance fluctuations\refnote{\cite{meso}}.
In these lectures we will discuss correlations in the bulk of Dirac spectra and
the microscopic spectral density.
The second type of application of Random Matrix Theory is as a schematic
model of disorder.  In this way one obtains $qualitative$ results which
may be helpful in understanding some physical phenomena. There are numerous
examples in this category. We only mention the Anderson 
model of localization\refnote{\cite{Anderson}}, neural 
networks\refnote{\cite{Sommers}},
the Gross-Witten model of QCD\refnote{\cite{GW}} 
and quantum gravity\refnote{\cite{Ginsparg}} . In these 
lectures we will discuss chiral random matrix models at nonzero 
temperature and chemical
potential. In particular, we will review recent work
by Stephanov\refnote{\cite{Stephanov}} on the quenched approximation at
nonzero chemical potential.

At nonzero chemical potential the QCD Dirac operator is nonhermitean with
eigenvalues scattered in the complex plane. As was first pointed out by
Fyodorov $et$ $al.$\refnote{\cite{fyodorovpoly}}, 
this leads to the possibility of 
a new type of universal behavior. Characteristic features of Dirac
spectra will be discussed at the end of this lecture. For a
review of nonhermitean random matrices, we refer to the talk by 
Nowak\refnote{\cite{Nowak}} in these proceedings.

In the first lecture we will review some general properties of Dirac spectra
including the Banks-Casher formula. From the zeros of the partition function
we will show that there is an intimate relation between chiral symmetry
breaking and correlations of Dirac eigenvalues. Starting from Leutwyler-Smilga
sum-rules the microscopic spectral density will be introduced. We will discuss
the statistical analysis of quantum spectra. It will be argued
that  spectral correlations of 'complex' systems 
are given by Random Matrix Theory.
We will end the first lecture with the introduction of 
chiral Random Matrix Theory.

In the second lecture we will compare the chiral random matrix model with 
QCD and discuss some of its properties. We will review recent
results that show that the microscopic
spectral density and eigenvalue correlations near zero virtuality are 
strongly universal. 
Lattice QCD results for the microscopic spectral
density and correlations in the bulk of the spectrum will be discussed in
detail. We will end the second lecture with a review  of chiral Random
Matrix Theory at nonzero chemical potential. New features of spectral
universality in nonhermitean matrices will be discussed.
 
\section{THE DIRAC SPECTRUM}
\subsection{Introduction}
The Euclidean QCD partition function is given by
\be
Z_{\rm QCD }(m,\theta) = 
\int dA \det{(\gamma D +m)} e^{-S_{\rm YM}/\hbar+i\theta\nu},
\label{QCDpart}
\ee
where $\gamma D=\gamma_\mu (\partial_\mu + i A_\mu)$ 
is the anti-Hermitean Dirac operator and $S_{\rm YM}$ is
the Yang-Mills action. The integral over field configurations includes a
sum over all topological sectors with topological charge $\nu$. Each sector
is weighted by $\exp(i\theta \nu)$. 
Phenomenologically the value of the vacuum
$\theta$-angle is consistent with zero.
We use the convention that the Euclidean gamma matrices are
Hermitean with $\{\gamma_\mu, \gamma_\nu\} = 2 \delta_{\mu\nu}$.
The integral is over all gauge field configurations,
and for definiteness, we assume a lattice regularization of the 
partition function. 

Our main object of interest is the spectrum of the Dirac operator. The
eigenvalues $\lambda_k$ are defined by
\be
\gamma D\phi_k = i\lambda_k \phi_k.
\ee
The spectral density is given by
\be
\rho(\lambda) = \sum_k \delta (\lambda-\lambda_k).
\ee
Correlations of the eigenvalues can be expressed in terms of 
the two-point correlation
function
\be
\rho_2(\lambda,\lambda') = \langle \rho(\lambda) \rho(\lambda') \rangle,
\ee
where $\langle \cdots \rangle$ denotes averaging with respect to the
QCD partition function (\ref{QCDpart}). The connected two-point
correlation function is obtained by subtraction of the product of the average
spectral densities
\be
\rho_c(\lambda,\lambda') = \rho_2(\lambda,\lambda')-
\langle \rho(\lambda)\rangle \langle \rho(\lambda') \rangle.
\ee

Because of the $U_A(1)$  symmetry
\be
\{\gamma_5, \gamma D\} = 0,
\ee
the eigenvalues occur in pairs $\pm \lambda$ or are zero. The eigenfunctions
are given by $\phi_k$ and $\gamma_5 \phi_k$, respectively. If $\gamma_5 \phi_k
= \pm \phi_k$, then necessarily $\lambda_k = 0$. This happens for a
solution of the Dirac operator in the field of an instanton. In
a sector with topological charge $\nu$ the Dirac operator has 
$\nu$ exact zero modes with
the same chirality. In order to represent the low energy sector of the 
Dirac operator for field configurations with topological charge $\nu$, 
it is natural to choose a chiral basis with 
$n$ right-handed states and $m\equiv n+\nu$ left-handed states. Then
the Dirac matrix has the block structure
\be
\left ( \begin{array}{cc} 0 & T \\ T^\dagger & 0 \end{array}\right ),
\label{block}
\ee
where $T$ is an $n\times m$ matrix. For $m=2$ and $n=1$, one can easily 
convince oneself that the Dirac matrix has exactly one zero eigenvalue. We 
leave it as an exercise to the reader to show that in general the Dirac matrix
has $|m-n|$ zero eigenvalues.

In terms of the eigenvalues of the Dirac operator, the QCD partition
function can be rewritten as
\be
Z_{\rm QCD }(m,\theta) =\sum_\nu e^{i\nu \theta}\prod_f m_f^{|\nu|} 
\int_{\nu} dA \prod_f\prod_{k}(\lambda_k^2 +m^2_f) e^{-S_{YM}/\hbar}
\label{ZQCDdet}
\ee
where $\int_\nu dA$ denotes the integral over field configurations with
topological charge $\nu$, and $\prod_f$ is the product
over $N_f$ flavors with mass $m_f$.
The partition function in the sector of topological charge $\nu$ is obtained
by Fourier inversion
\be
Z_\nu(m) = \frac 1{2\pi}\int_0^{2\pi} d\theta e^{-i\nu\theta} 
Z_{\rm QCD}(m,\theta).
\label{znum}
\ee 
The fluctuations of the eigenvalues of the QCD Dirac operator 
are induced by the fluctuations of the gauge fields. 
Formally, one can think of integrating out all gauge field configurations
for fixed values of the Dirac eigenvalues. 
The transformation of integration variables from the fields, $A$, to the
eigenvalues, $\lambda_k$,  leads to a nontrivial "Jacobian".
Universality in Dirac spectra has its origin in this "Jacobian".

\begin{center}
\begin{figure}[!ht]
\centering\includegraphics[width=100mm,angle=0]
{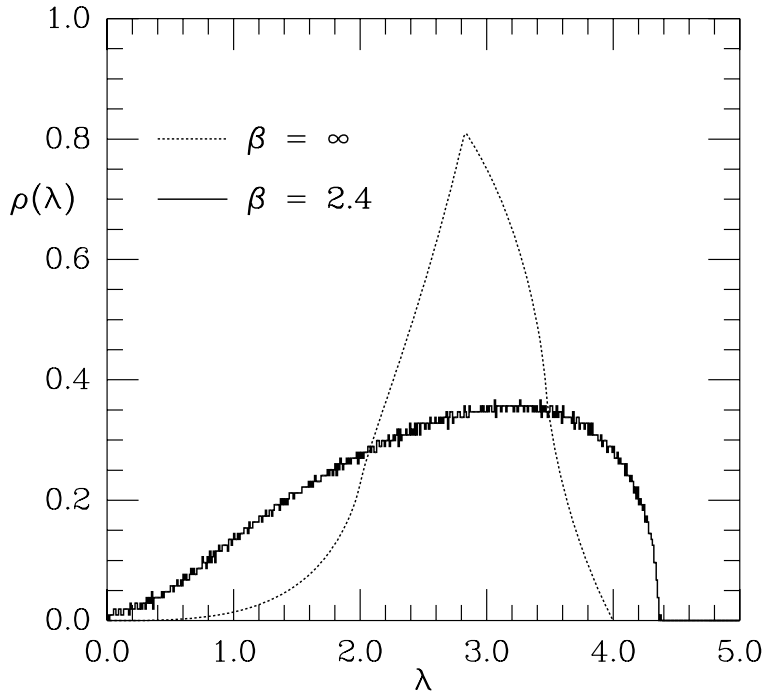}
\caption{The free Dirac spectral density (dotted curve) and the spectral
density of the Kogut-Susskind Dirac operator
for a gauge field configuration generated with $\beta = 2.4$
(histogram). Both spectral densities are on a $12^4$ lattice
and are normalized to unit area.} 
\label{fig1}
\vspace*{-0.5cm}
\end{figure}
\end{center}

The free Dirac spectrum can be obtained immediately from the square of the
Dirac operator. For a box of volume $L_1\times 
L_2\times L_3\times L_4$ one finds
\be
\lambda_{\vec n} = 2\pi\left (
(\frac {n_1}{L_1})^2+(\frac {n_2}{L_2})^2+
(\frac {n_3}{L_3})^2+(\frac {n_4+0.5}{L_4})^2\right)^{1/2},
\ee
where we have used periodic boundary conditions in the spatial directions
and anti-periodic boundary conditions in the time direction.
The spectral density is obtained by counting the total number of eigenvalues
in a sphere of radius $\lambda L/2\pi$. The result is
\be
\rho^{\rm free}(\lambda) \sim V \lambda^3.
\label{rhofree}
\ee
For future reference, we note that in the generic case, when the sides of
the hypercube are related by an irrational number, the eigenvalues are 
uncorrelated, i.e.
\be
\rho_2(\lambda, \lambda') = \langle \rho(\lambda)\rangle 
\langle \rho(\lambda')\rangle.
\ee

Two examples of Dirac spectra are shown in Fig. 1. The dotted curve 
represents the free Kogut-Susskind Dirac spectrum  on a $12^4$ lattice
with periodic boundary conditions in the spatial directions
anti-periodic boundary conditions in the time direction. For an
$N_1\times N_2\times N_3 \times N_4$ lattice this spectrum is given by
\be
\lambda_{\vec n} = 2\left ( 
\sin^2 \left ( \frac {\pi n_1}{N_1} \right ) +
\sin^2 \left ( \frac {\pi n_2}{N_2} \right ) +
\sin^2 \left ( \frac {\pi n_3}{N_3} \right ) +
\sin^2 \left ( \frac {\pi (n_4+0.5)}{N_4} \right ) \right)^{1/2}.
\ee
Here, $n_i = 0, 1, \cdots, [N_i/2]$ ($i = 1,\, 2, \,3$) and $n_4 = 0, 1, \cdots,
[(N_4+1)/2]$. The  Kogut-Susskind Dirac
spectrum for an $SU(2)$ gauge field configuration with $\beta =2.4$
on the same size lattice is shown by the histogram in the same figure (full 
curve). We clearly observe
an accumulation of small eigenvalues.

\subsection{The Banks-Casher Relation}

The order parameter of the chiral phase transition,
$\langle \bar \psi \psi \rangle$,
is nonzero only below the critical temperature. 
As was shown by Banks and Casher\refnote{\cite{Banks-Casher}}, 
$\langle \bar \psi \psi \rangle$ is directly related to the eigenvalue density
of the QCD Dirac operator per unit four-volume 
\be
\Sigma \equiv 
|\langle \bar \psi \psi \rangle| =\lim\frac {\pi \langle {\rho(0)}\rangle}V.
\label{bankscasher}
\ee
It is elementary to derive this relation.
The chiral condensate follows from
the partition function (\ref{ZQCDdet}) (all quark mass are chosen equal),
\be
\langle \bar \psi \psi \rangle  = -\lim\frac 1{VN_f}\partial_m \log Z(m) 
=-\lim\frac 1V \langle \sum_k \frac {2m}{\lambda_k^2 + m^2}\rangle.
\label{preBC}
\ee
If we express the sum as an integral over the average spectral density,
and take the thermodynamic limit before the chiral limit, so that many
eigenvalues are less than $m$, we recover (\ref{bankscasher}). The order of the
limits in (\ref{bankscasher}) is important. First we take the 
thermodynamic limit, next the chiral limit and, finally, the field theory
limit. As can be observed from (\ref{preBC}) the
sign of $\langle \bar \psi \psi \rangle $ changes if $m$ crosses the 
imaginary axis. 

An important consequence of the Bank-Casher formula (\ref{bankscasher})
is that the eigenvalues near zero virtuality are spaced as  
\be
\Delta \lambda = 1/{\rho(0)} = {\pi}/{\Sigma V}.
\ee
This should be contrasted with the eigenvalue spectrum 
of the non-interacting 
Dirac operator. Then one obtains from (\ref{rhofree})
an eigenvalue spacing equal to  $\Delta \lambda \sim 1/V^{1/4}$.
Clearly, the presence of gauge fields leads to a strong modification of
the spectrum near zero virtuality. Strong interactions result in the 
coupling of many degrees of freedom leading to extended states and correlated
eigenvalues.
On the other hand, for uncorrelated eigenvalues, the eigenvalue distribution
factorizes, and for $\lambda \ne 0$,
we have $\rho(\lambda) \sim \lambda^{2N_f+1}$ in the chiral limit, 
i.e. no breaking of chiral symmetry. 
One consequence of the interactions is level repulsion
of neighboring eigenvalues. Therefore, the two smallest eigenvalues of
the Dirac operator, $\pm \lambda_{\rm min}$ repel each other, and the Dirac
spectrum will have a gap at $\lambda = 0$ with a width of the order of 
$1/\Sigma V$.  

\section{Spectral Correlations and Zeros of the Partition Function}

The study of zeros of the partition function has been a fruitful
tool in statistical mechanics\refnote{\cite{Frank,Shrock}}. In QCD,
both zeros in the in the complex fugacity plane and the complex mass
plane have been studied\refnote{\cite{vink,barbourqed}}.
Since the QCD partition function is a polynomial in $m$ it can be factorized
as (all quark masses are taken to be equal to $m$)
\be
Z_{\rm QCD}(m,\theta) = \prod_k (m-m_k).
\ee
Because configurations of opposite topological charge occur with the same 
probability,
the coefficients of this polynomial are real, and 
the zeros appear in complex conjugate pairs.  For an even number of flavors
the zeros occur in pairs $\pm m_k$. In a sector with topological charge
$\nu$, this is also the case for even
$N_f \times \nu$. The chiral condensate is given by
\be
\Sigma(m) = -\lim\frac 1{VN_f} \partial_m \log Z_{\rm QCD}(m,\theta)=
-\lim\frac 1{VN_f}\sum_k \frac 1{m-m_k}.
\ee
For an even number of flavors, $\Sigma(m)$ is an odd function of $m$.
In order to have a discontinuity at $m=0$, the zeros in this region
have to coalesce into a 
cut along the imaginary axis in the thermodynamic limit.

In the hypothetical case that the eigenvalues of the Dirac operator do not
fluctuate the zeros are located at $m_k = \pm i\lambda_k$. In the opposite
case, of uncorrelated eigenvalues, the eigenvalue distribution factorizes
and all zeros are located at $\pm i \sigma$, where $\sigma^2 = \langle
\lambda^2_k\rangle$.  
As a result, the chiral condensate does not 
show a discontinuity across the imaginary axis and is equal to zero.

We hope to convince the reader that the presence of a  
discontinuity is intimately related to  correlations of 
eigenvalues\refnote{\cite{oscor}}. 
Let us study the effect of pair correlations for one flavor in 
the sector of zero topological charge. 
The fermion determinant can be written as
\be
\langle \prod_k (m^2+\lambda_k^2) \rangle = \sum_k 
\left (\begin{array}{c}  N \\ k \end{array} \right ) 
m^{2(N-k)}\langle  \lambda_1^2 \cdots \lambda_k^2\rangle.
\ee
There are $$\left (\begin{array}{c}  k \\ 2l \end{array} \right ) 
(2l-1)!!$$ ways of selecting $l$ pairs from $\lambda_1^2 \cdots \lambda_k^2$.
The average of each pair of different eigenvalues is given by
\be
\langle \lambda_m^2 \lambda_n^2 \rangle =
\sigma^4 +C_2,
\ee
where $\sigma^2$ is the expectation value of $\lambda_k^2$ and $C_2$
is the connected correlator
\be
C_2 = \langle \lambda_m^2 \lambda_n^2 \rangle - 
\langle \lambda_m^2 \rangle \langle \lambda_n^2 \rangle, \qquad m\ne n.
\ee
This results in the partition function
\be
Z(m) = \sum_{k=0}^N \sum_{l=0}^{[\frac k2 ]} m^{2(N-k)} 
\left (\begin{array}{c}  N \\ k \end{array} \right ) 
\left (\begin{array}{c}  k \\ 2l \end{array} \right ) 
(2l-1)!! C_2^l \sigma^{2(k-2l)}.
\ee
After interchanging the two sums, one can easily show that $Z(m)$ can be
expressed as a multiple of a Hermite polynomial
\be
Z(m) = (-\frac{C_2}2)^{N/2} {\rm H}_N((\sigma^2+m^2)/\sqrt{-2C_2}).
\ee
In terms of the zeros of the Hermite polynomials, $z_k$, the zeros of
the partition function are located at
\be
m_k^2 = z_k \sqrt{-2C_2} -\sigma^2.
\ee
Asymptotically, the zeros of the Hermite polynomials are given by
$z_k \approx \pi k/2\sqrt N$.
In order for the zeros to join into a cut in the thermodynamic limit, they have
to be spaced as $ \sim 1/N$.  This requires that 
\be
C_2 \sim -\frac 1N.
\ee 
The density of zeros is then given by
\be
\frac{dk}{dm} \sim N m.
\ee
We conclude that pair correlations are sufficient to generate a cut of
$Z(m)$ in the complex $m$-plane, but the chiral symmetry remains unbroken.
Pair correlations alone cannot suppress the effect of the fermion 
determinant.

\subsection{Leutwyler-Smilga Sum Rules}
We have shown that pair-correlations are not sufficient to generate
a discontinuity in the chiral condensate. In this subsection we 
start from the assumption that
chiral symmetry is broken spontaneously, and look for consistency 
conditions this imposes
on the Dirac spectrum. As has been 
argued by Gasser and Leutwyler\refnote{\cite{GL}} and 
Leutwyler and Smilga\refnote{\cite{LS}}, in the mesoscopic range (\ref{range}),
the  mass dependence of the QCD partition function is given by 
(for simplicity, all quark masses have been taken equal)
\be
Z^{\rm eff}(m,\theta) 
\sim \int_{U\in G/H} dU e^{mV\Sigma {\rm Re\,Tr}\, Ue^{i\theta/N_f}}.
\label{zm}
\ee
The integral is over the Goldstone manifold associated with chiral
symmetry breaking from $G$ to $H$. For three or more colors with
fundamental fermions $G/H = SU(N_f) \times SU(N_f)/SU(N_f)$.
The finite volume 
partition function in the sector of topological charge $\nu$ follows
by Fourier inversion according to (\ref{znum}).
The partition function for $\nu = 0$ is thus given by (\ref{zm})
with the integration over $SU(N_f)$ replaced by an integral over
$U(N_f)$. The case of $N_f= 1$ is particularly simple. 
Then only a $U(1)$ integration remains, and 
the partition function is given by\refnote{\cite{LS}}
$Z_{\nu = 0}^{\rm eff}(m) = I_0(mV\Sigma)$.
Its zeros are regularly spaced along 
the imaginary axis in the complex $m$-plane, and, in the thermodynamic limit, 
they coalesce into a cut.

The Leutwyler-Smilga sum-rules are obtained by expanding the partition
function $Z_\nu(m)$ in powers of $m$ before and after 
averaging over the gauge field configurations and equating the 
coefficients. This corresponds to an expansion in powers of $m$ of
both the QCD partition function (\ref{QCDpart}) and the finite
volume partition function (\ref{zm}) in the sector
of topological charge $\nu$.
As an example, 
we consider the coefficients of $m^2$ in the 
sector with $\nu = 0$. This results in the sum-rule
\be
\langle {\sum}' \frac 1{\lambda_k^2}\rangle  = \frac {\Sigma^2 V^2}{4N_f},
\label{LS2}
\ee
where the prime indicates that the sum is restricted to nonzero positive 
eigenvalues.

The next order sum rules are obtained by equating the coefficients of order 
$m^4$. They can be combined into
\be
\langle {\sum_{k,l}}' \frac 1{\lambda_k^2 \lambda_l^2}\rangle  
-\langle {\sum_k}' \frac 1{\lambda_k^2}\rangle \langle  {\sum_l}' 
\frac 1{\lambda_l^2}
\rangle
= \frac {\Sigma^4 V^4}{16N_f^2(N_f^2 -1)}.
\label{LSpair}
\ee
We conclude that chiral symmetry breaking leads to correlations
of the inverse eigenvalues. However, if one performs an analysis similar 
to the one in previous section, it can be shown easily
that pair correlations given by (\ref{LSpair}) do not result
in a cut in the complex $m$-plane. Apparently, chiral symmetry breaking 
requires a subtle interplay of all types of correlations.  

For two colors with fundamental fermions or for adjoint fermions the
pattern of chiral symmetry breaking is different. 
Sum rules for the inverse eigenvalues can be derived 
along the same lines. The general expression for the simplest sum-rule
can be summarized as\refnote{\cite{SmV,HVeff}}
\be
\langle {\sum}' \frac 1{\lambda_k^2}\rangle  = \frac {\Sigma^2 V^2}
{4(|\nu| +({\rm dim}(G/H) + 1)/N_f)}.
\ee

The Leutwyler-Smilga sum-rules can be expressed as an integral over the average
spectral density and spectral correlation functions. For the sum rule
(\ref{LS2}) this results in
\be
\frac 1{V^2 \Sigma^2} \int \frac {\langle\rho(\lambda)\rangle
d\lambda}{\lambda^2} = \frac 
1{4N_f}.
\ee
If we introduce the microscopic variable 
\be
u = \lambda V \Sigma,
\ee
this integral can be rewritten as
\be
\int \frac 1{V\Sigma} \langle\rho(\frac u{V\Sigma}) \rangle
\frac {du}{u^2} = \frac 1{4N_f}.
\ee
The thermodynamic limit of the combination
that enters in the integrand, 
\be
\rho_S(u) = \lim_{V\rightarrow \infty} \frac 1{V\Sigma} \langle
\rho(\frac u{V\Sigma})\rangle,
\label{rhosu}
\ee
will be called the microscopic
spectral density\refnote{\cite{SVR}}. This limit exists if chiral symmetry is
broken. Our
conjecture is that $\rho_S(u)$ is a universal function that only depends
on the global symmetries of the QCD partition function. Because of universality
it can be derived from the simplest theory with the global 
symmetries of the QCD
partition function. Such theory is a chiral Random Matrix Theory which
will be introduced later in these lectures.

We emphasize again that the
$U_A(1)$ symmetry of the QCD Dirac spectrum leads to two 
different types of eigenvalue correlations: spectral
correlations in the bulk of the spectrum and spectral correlations near zero 
virtuality. The simplest example of correlations of the latter type is
the microscopic spectral density defined in (\ref{rhosu}).

We close this subsection with two unrelated side remarks. First,  the QCD 
Dirac operator 
is only determined up to a constant matrix. We can exploit this freedom to 
obtain a Dirac operator that is maximally symmetric. 
For example, the Wilson lattice QCD Dirac operator, $D^W$, is neither Hermitean
nor anti-Hermitean, but $\gamma_5 D^W$ is Hermitean.

Second, the QCD partition function 
can be expanded in powers of $m^2$ 
before or after averaging over the gauge field configurations. In the latter
case one obtains sum rules for the inverse zeros of the partition function.
As an example we quote,
\be
\left . \sum \frac 1{m_k^2} \right |_{\nu = 0}= \frac {\Sigma^2 V^2}4,
\ee
where we have averaged over field configurations with zero topological charge.

\section{SPECTRAL CORRELATIONS IN COMPLEX SYSTEMS}

\subsection{Statistical Analysis of Spectra}
Spectra for a wide range of complex quantum systems 
have been studied both experimentally and numerically (a excellent 
recent review has been given by Guhr, M\"uller-Groeling and 
Weidenm\"uller\refnote{\cite{hdgang}}).
One basic observation
is that the scale of variations of the average spectral
density and the scale of the spectral fluctuations separate. 
This allows us to unfold the spectrum, i.e. we rescale the 
spectrum in units of the local average level spacing. 
Specifically, the unfolded spectrum is given by
\be
\lambda_k^{\rm unf} = \int_{\infty}^{\lambda_k} \langle \rho(\lambda')\rangle
d \lambda',
\ee
with unfolded spectral density 
\be
\rho_{\rm unf} (\lambda) = \sum_k \delta(\lambda -\lambda_k^{\rm unf}).
\ee

The fluctuations of the
unfolded spectrum can be measured by suitable statistics. We will consider the
nearest neighbor spacing distribution, $P(S)$, and moments of the number of
levels in an interval containing $n$ levels on average. In particular, we
will consider the
number variance, $\Sigma_2(n)$, and the first two cumulants, $\gamma_1(n)$ and
$\gamma_2(n)$. Another useful statistic is the
$\Delta_3(n)$-statistic introduced by Dyson and Mehta\refnote{\cite{deltaDM}}.
It is related to $\Sigma_2(n)$
via a smoothening kernel. The advantage of this statistic is that its 
fluctuations as a function of $n$ are greatly reduced. 
Both $\Sigma_2(n)$ and $\Delta_3(n)$ can be obtained from the pair correlation
function defined as
\be
Y_2(\lambda,\lambda') =-\langle{\rho_{\rm unf}(\lambda)
\rho_{\rm unf}(\lambda')} \rangle+
\langle {\rho_{\rm unf}(\lambda)}\rangle \langle {\rho_{\rm unf}(\lambda')}
\rangle
\ee

Analytical expressions for the above  statistics can be obtained for
the eigenvalues of the invariant random matrix ensembles. 
They are defined as ensembles of Hermitean 
matrices with Gaussian independently distributed matrix elements, i.e.
with probability distribution given by
\be
P(H) \sim e^{-\frac {N\beta} 2 {\rm Tr } H^\dagger H}.
\label{phinv}
\ee 
Depending on the anti-unitary symmetry, the matrix elements are real, complex
or quaternion real. They are called the Gaussian Orthogonal Ensemble (GOE),
the Gaussian Unitary Ensemble (GUE) and the Gaussian Symplectic Ensemble
(GSE), respectively. Each ensemble is characterized by its Dyson index
$\beta$ which is defined as the number of independent variables per matrix
element. For the GOE, GUE and the GSE we thus have $\beta =1, \, 2$ and 4,
respectively. 

Independent of the value of $\beta$, the average spectral
density is the semicircle,
\be
\langle \rho(\lambda)\rangle = \frac N \pi \sqrt{2-\lambda^2}.
\ee

Analytical results for all spectral correlation functions 
have been derived for each of the three ensembles\refnote{\cite{Mehta}} 
via the orthogonal polynomial method. 
We only quote the most important results.
The nearest neighbor spacing distribution, which is known exactly in terms of a
power series, is well approximated by
\be
P(S) \sim S^{\beta}\exp(-a_\beta S^2),
\ee 
where $a_\beta$ is a constant of order one.
The asymptotic behaviour of the pair correlation functions is given
by\refnote{\cite{Mehta}}
\be
Y_2(\lambda,\lambda') &\sim& \frac {1}{\pi^2(\lambda-\lambda')^2} 
\quad {\rm for} \quad \beta = 1,\\
Y_2(\lambda,\lambda') &\sim& \frac {\sin^2\pi(\lambda 
-\lambda')}{\pi^2(\lambda-\lambda')^2} \quad {\rm for} \quad \beta = 2,\\
Y_2(\lambda,\lambda') &\sim& -\frac{\cos 
2\pi(\lambda-\lambda')}{4(\lambda-\lambda')} +\frac {1+ (\pi/2)\sin2\pi(\lambda 
-\lambda')}{4\pi^2(\lambda-\lambda')^2} \quad {\rm for} \quad \beta = 4.
\ee
The $1/(\lambda-\lambda')^2$ tail of the pair correlation function results
in a logarithmic dependence of the asymptotic behavior of $\Sigma_2(n)$
and $\Delta_3(n)$,
\be
\Sigma_2(n) \sim (2/\pi^2\beta) \log n \quad {\rm and} \quad  
\Delta_3(n) \sim \beta \Sigma_2(n)/2.
\ee
Characteristic features of random matrix correlations are
level repulsion at short distances and a strong suppression
of fluctuations at large distances.

For uncorrelated eigenvalues the level repulsion is absent and
one finds 
\be
P(S) = \exp(-S),
\ee
and 
\be
\Sigma_2(n) = n  \quad {\rm and} \quad \Delta_3(n) = n/15.
\ee

\subsection{Spectral Universality}

The main conclusion of numerous studies of eigenvalue spectra 
of complex systems is that
spectral correlations of classically chaotic systems are given 
by RMT\refnote{\cite{hdgang}}. 
As illustration we mention three examples from completely
different fields. The first example is the nuclear data ensemble in which
the above statistics are evaluated by superimposing level spectra of
many different nuclei\refnote{\cite{Haq}}. The second example concerns 
correlations of acoustic resonances in irregular 
quartz blocks\refnote{\cite{Guhr}}. In both cases the statistics that 
were considered are,
within experimental accuracy, in complete agreement with
the GOE statistics. The third example
pertains to the zeros of Riemann's zeta function. Extensive numerical
calculations\refnote{\cite{Odlyzko}} have shown that asymptotically, for large
imaginary part, the correlations of the zeros are given by the GUE.

The Gaussian random matrix ensembles introduced above
can be obtained\refnote{\cite{Mehta}} from two assumptions:
i) The probability distribution is invariant under unitarity
transformations;   ii) The matrix elements are statistically independent.
If the invariance assumption is dropped it can be shown
with the theory of free random 
variables\refnote{\cite{Voiculescu}} 
that the average spectral density is still given by a semicircle
if the variance of the probability distribution is finite. For example,
if the matrix elements are distributed according to a rectangular 
distribution, the average spectral density is a semicircle.
On the other
hand, if the independence assumption is released the average spectral
density is typically not a semicircle. For example, this is the case
if the  quadratic potential in the probability distribution is replaced
by a more complicated polynomial potential $V(H)$. Using the supersymmetric
method for Random Matrix Theory, it was shown by 
Hackenbroich and Weidenm\"uller\refnote{\cite{Hack}} that 
the same supersymmetric
nonlinear $\sigma$-model is obtained for a wide range of potentials $V(H)$.
This implies that spectral
correlations of the unfolded eigenvalues are independent of the
potential. Remarkably, this result could
be proved for all three values of the Dyson index. 

An explicit construction 
of the correlation functions using orthogonal polynomials could only 
be performed\refnote{\cite{kanzieper,Eynard,Zee-poly,Deo}} for $\beta = 2$.  
Several examples have been considered where both the invariance assumption
and the independence assumption are relaxed. We mention $H \rightarrow
H+A$, where $A$ is an arbitrary fixed matrix, and
the probability distribution of $H$ is given by a polynomial 
$V(H)$. It was shown
by P. Zinn-Justin\refnote{\cite{Pzinn}} 
that also in this case the spectral correlations are given
by the invariant random matrix ensembles.  For a Gaussian probability
distribution the proof was given 
by Br\'ezin and Hikami\refnote{\cite{Brezin-Hikami}}.

The domain of universality has been extended 
in the direction of real physical systems by means of 
the Gaussian embedded ensembles\refnote{\cite{french,zirntwo}}. The 
simplest example is the ensemble of matrix elements of $n$-particle Slater
determinants of a two-body operator with random two-particle matrix 
elements. It can be shown analytically 
that the average spectral density is a Gaussian.
However, according to substantial numerical evidence, the spectral correlations
are in complete agreement with the invariant random matrix 
ensembles\refnote{\cite{french}}.

Universal spectral correlations are obtained in the thermodynamic (or 
semi-classical) limit, $N\rightarrow \infty$,  with $(E-E')N$ fixed. 
Alternatively, one can take the thermodynamic limit with  $E-E'$ fixed.
This leads to the Ambjorn-Jurkiewicz-Makeenko 
(AJM-~)universality\refnote{\cite{AJM}} for
smoothened correlation functions\refnote{\cite{Zee-poly}}. They are obtained
from the exact correlation functions by replacing the oscillating terms by
their average over a scale much larger than $1/N$, but small compared to the
secular variation of the average spectral density. 
The result for the 
two-point correlation function is given by    
a smoothening of the leading order asymptotic result\refnote{\cite{Zee-poly}}. 
However, correlations at macroscopic distance in energy
should include another ingredient. Namely, correlations 
resulting from the compactness of the
support of the spectrum. 
Indeed, Ambjorn, 
Jurkiewicz and Makeenko showed\refnote{\cite{AJM}} that the smoothened 
correlation functions for an arbitrary potential are determined 
completely by
the endpoints of the spectrum. This theorem was proved 
for all three ensembles by 
Beenakker\refnote{\cite{Been}} and it was generalized to multi-cut
potentials by Akemann and
Ambjorn\refnote{\cite{Akemann}}. A general expression for 
the smoothened correlation
function in terms of one-point Greens functions has been derived
for different deformations of the invariant random matrix 
ensembles\refnote{\cite{french,us,bzgeneral}}.
An interesting question is whether correlations given by
this general expression are also found in real physical systems. 
There are several indications that the answer to this question is 
negative. First of all, AJM universality is
closely related to the compactness of the support of the spectrum.
In real physical systems, the spectral density usually increases with
energy so that the average resolvent does not even exist (only
the difference of two resolvents enters in the general expression). 
Second, in fully
chaotic systems, with microscopic correlations given by Random Matrix Theory,
it was shown by Berry\refnote{\cite{Berry}} that 
the asymptotics of the two-point correlation function, as 
measured by the $\Delta_3$-statistic, is determined by the shortest periodic
trajectory.

Smoothened correlators are obtained via a perturbative expansion of the
correlation functions.
In Random Matrix Theory this is equivalent to a loop expansion in $1/N$.
The full non-perturbative result cannot be obtained this way. That requires
the use of orthogonal polynomials or the super-symmetric method of Random
Matrix Theory. 

Another type of universal behavior is given by correlations
in the neighborhood of the largest
eigenvalue. It was shown by Kanzieper and Freilikher\refnote{\cite{KanFrei}}
that for an arbitrary potential the spectral
correlations at the soft edge of the spectrum are given by the 
Airy kernel. However, we are not aware of any physical applications of such
correlations. Certainly, this type of universality is restricted to systems
with a resolvent that is defined by a finite cuts
on the real axis. For example, for the embedded ensembles with a
Gaussian spectral density, we do not expect
such type of  universality. 

We have seen a large number of examples that fall into one of
universality classes of the invariant random matrix ensembles. This calls 
out for a more general approach. 
Naturally, one thinks in terms of the renormalization
group. This approach was pioneered by Br\'ezin and 
Zinn-Justin\refnote{\cite{Brezinn}}. The idea is to integrate out rows and 
columns of a random matrix and to show that the Gaussian ensembles
are  a stable fixed point. This was made more explicit in a paper
by Higuchi {\it et al.}\refnote{\cite{Nishigaki-unir}}.
However, much more work is required 
to arrive at a natural proof of spectral universality.

Although the above mentioned universality studies provide support for
the validity of the Bohigas conjecture, the ultimate goal is to derive it
directly from the underlying classical dynamics. 
An important first step in this direction
was made by Berry\refnote{\cite{Berry}}. He showed that the asymptotics 
of the two-point
correlation function is related to sum-rules for isolated
classical trajectories. Another interesting approach was introduced by
Andreev {\it et al.}\refnote{\cite{andreev}} who 
where able to obtain a supersymmetric nonlinear sigma model from spectral
averaging. In this context we also mention the work of Altland and 
Zirnbauer\refnote{\cite{Martinkick}}
who showed that the kicked rotor can be mapped onto a supersymmetric
sigma model.

\section{CHIRAL RANDOM MATRIX THEORY}
\subsection{Introduction of the Model}
In this section we will introduce an instanton 
liquid\refnote{\cite{shurrev,diakonov}} inspired 
chiral RMT for the QCD partition function. 
In the spirit of the invariant random matrix ensembles 
we construct a model for the Dirac
operator with the global symmetries of the QCD partition function as input, but
otherwise  Gaussian random matrix elements. 
The chRMT that obeys these conditions is defined 
by\refnote{\cite{SVR,V,VZ,Vlattice}}
\be
Z_{N_f,\nu}^\beta(m_1,\cdots, m_{N_f}) = 
\int DW \prod_{f= 1}^{N_f} \det({\rm \cal D} +m_f)
e^{-\frac{N\Sigma^2 \beta}4 {\rm Tr}W^\dagger W},
\label{zrandom}
\ee
where
\be
{\cal D} = \left (\begin{array}{cc} 0 & iW\\
iW^\dagger & 0 \end{array} \right ),
\label{diracop}
\ee
and $W$ is a $n\times m$ matrix with $\nu = |n-m|$ and
$N= n+m$. The matrix elements of $W$ are either real ($\beta = 1$, chiral
Gaussian Orthogonal Ensemble (chGOE)), complex
($\beta = 2$, chiral Gaussian Unitary Ensemble (chGUE)),
or quaternion real ($\beta = 4$, chiral Gaussian Symplectic Ensemble (chGSE)).
As is the case in QCD, we assume that $\nu$ does not exceed $\sqrt N$, so that,
to a good approximation, $n = N/2$.

This model reproduces the following symmetries of the QCD partition
function:

\begin{itemize}
\item The $U_A(1)$ symmetry. All nonzero eigenvalues of the random matrix
Dirac operator occur in pairs $\pm \lambda$ or are zero.

\item  The topological structure of the QCD partition function. The 
Dirac matrix has exactly $|\nu|\equiv |n-m|$ zero eigenvalues. This identifies
$\nu$ as the topological sector of the model.

\item The flavor symmetry is the same as in QCD. For $\beta = 2$ it is
$SU(N_f) \times SU(N_f)$, for $\beta = 1$ it is $SU(2N_f)$ and for
$\beta = 4$ it is $SU(N_f)$.

\item The chiral symmetry is broken spontaneously with 
chiral condensate given by
\be                                                      
\Sigma = \lim_{N\rightarrow \infty} {\pi \rho(0)}/N.
\ee
($N$ is interpreted as the (dimensionless) volume of space
time.) The symmetry breaking pattern
is\refnote{\cite{SmV}} $SU(N_f) \times SU(N_f)/SU(N_f)$,
$SU(2N_f)/Sp(N_f)$ and $SU(N_f)/O(N_f)$ for $\beta = 2$, 1 and 4,
respectively, the same as in QCD\refnote{\cite{Shifman-three}}.

\item The anti-unitary symmetries. 
For three or more colors with
fundamental fermions the Dirac operator has no anti-unitary symmetries,
and generically, the matrix elements of the Dirac operator are
complex. The matrix elements $W_{kl}$ of the corresponding random matrix
ensemble are chosen arbitrary complex as well ($\beta =2$).
For $N_c =2$,  the Dirac operator in the fundamental representation
satisfies
\be
[C\tau_2 K, i\gamma D] = 0,
\ee
where $C$ is the charge conjugation matrix and $K$ is the complex conjugation
operator.
Because, $(C\tau_2 K)^2 =1$, the matrix elements of the Dirac operator
can always be chosen real, and the corresponding random matrix ensemble
is defined with real matrix elements ($\beta = 1$).
For two or more colors with gauge fields
in the adjoint representation the anti-unitary symmetry of the Dirac operator
is given by
\be
[CK, i\gamma D] =0.
\ee
Because $(CK)^2 = -1$, it is possible to rearrange the matrix elements of
the Dirac operator into real quaternions. The matrix elements $W_{kl}$ of the
corresponding random matrix ensemble are chosen quaternion real 
$(\beta =4)$.
\end{itemize}

Together with the invariant random matrix ensembles, the chiral ensembles are
part of a larger classification scheme. 
Apart from the random matrix ensembles discussed in this review, this 
classification also includes random matrix models for disordered
super-conductors\refnote{\cite{Altland}}.
As pointed out by 
Zirnbauer\refnote{\cite{class}}, 
all known universality classes 
of Hermitean 
random matrices are tangent to the large classes of 
symmetric spaces in the classification given by Cartan. 
There is a 
one-to-one correspondence between this classification and the 
classification of the large families of Riemannian symmetric 
superspaces\refnote{\cite{class}}.

\subsection{Selected Results for the Chiral Random Matrix Ensembles}

The joint eigenvalue distribution follows from (\ref{zrandom}) by choosing the
eigenvalues and eigenvectors as new integration variables. 
For $ N_f$ flavors and topological charge $\nu$  it is given 
by\refnote{\cite{V}}
\be
\rho_\beta(\lambda_1, \cdots, \lambda_n) = C_{\beta, n}
\prod_{k<l} |\lambda_k^2 -\lambda_l^2|^\beta \prod_{k}
\lambda_k^{2N_f +\beta\nu+\beta -1}
\exp\left ({-\frac{n\beta\Sigma^2}{2} \sum_k \lambda^2_k}\right ),
\label{rhototal}
\ee
where $C_{\beta,n}$ is a normalization constant and $\nu \ge 0$. 
For $\beta = 2$ the average
spectral density and the spectral correlation functions can be derived
from (\ref{rhototal}) with the help of the orthogonal polynomial 
method\refnote{\cite{Mehta}}. The associated polynomials are
the generalized Laguerre polynomials. That is why this ensemble  
is also known as the
Laguerre ensemble\refnote{\cite{Kahn,BRONK}}. The spectral density 
and the two-point 
correlation function were also derived within the framework
of the supersymmetric method of Random Matrix Theory\refnote{\cite{ast}}. 
The calculation of the average spectral density and the spectral
correlations functions for $\beta =1$ and
$\beta = 4$ is much more complicated. However, with the help of 
skew-orthogonal polynomials\refnote{\cite{Dyson-skew,Mehtaskew,nagao}} exact
analytical results for finite $N$ can be obtained as well. 

From the properties of the Laguerre polynomials it can be shown 
that, independent of the value of $\beta$, 
the average spectral density is a semi-circle
\be
\rho(\lambda) = (n\Sigma^2/\pi)\sqrt{4/\Sigma^2-\lambda^2}.
\ee

The microscopic spectral density can be derived from the limit
(\ref{rhosu}) of the exact spectral density for finite $N$.
For $N_c = 3$, $N_f$ flavors and topological charge $\nu$
it is given by\refnote{\cite{V}}
\be
\rho_S(z) = \frac z2 \left ( J^2_{a}(z) -
J_{a+1}(z)J_{a-1}(z)\right),
\label{micro2}
\ee
where $a = N_f + |\nu|$. 
The expressions for  $SU(2)$ with fundamental fermions ($\beta =1$) are much
more complicated. In this case we find
the microscopic spectral density\refnote{\cite{V2}}
\be
\rho_S(z) = \frac {\Sigma}{4} J_{2a+1}(z{\Sigma}) &+& \frac {\Sigma}{2}
\int_0^\infty dw (zw)^{2a+1} \epsilon(z-w)
\left ( \frac 1w \frac d{dw} - \frac 1z \frac d{dz}\right )\nonumber \\
&\times&
\frac{wJ_{2a}(z{\Sigma})J_{2a-1}( w{\Sigma})
-zJ_{2a-1}(z{\Sigma})J_{2a}(w{\Sigma})}{(zw)^{2a}(z^2-w^2)},
\label{micro1}
\ee
where $a$ is the combination
\be
a = N_f -\frac 12 +\frac {\nu}2.
\label{5.8}
\ee
The microscopic spectral density in the
symmetry class with $\beta = 4$ was first calculated by Nagao and 
Forrester\refnote{\cite{nagao-forrester}}. It is given by
\begin{eqnarray}
  \label{eq3}
  \rho_S(z)=2 z^2\int_0^1duu^2\int_0^1dv&&[J_{4a-1}(2uvz)J_{4a}(2uz)
  -vJ_{4a-1}(2uz)J_{4a}(2uvz)]
\label{micro4}
\end{eqnarray}
with $4a=N_f+2|\nu|+1$.  

The spectral correlations in the bulk of the spectrum are given 
by the invariant random matrix ensemble with the same value
of $\beta$. For $\beta =2$ this was already shown three decades ago
by Fox and Kahn\refnote{\cite{Kahn}}. For $\beta =1$ and $\beta =4 $ this
was only proved recently\refnote{\cite{nagao}}.       

\subsection{Duality between Flavor and Topology}

As one can observe from the joint eigenvalue distribution, for $\beta = 2$
the dependence on $N_f$ and $\nu$ enters only through the combination
$N_f + \nu$. This allows for the possibility of trading topology for
flavors. In this section we will work out this duality for the
finite volume partition function. This relation completes the proof
of the conjectured expression\refnote{\cite{Sener3}} for the finite 
volume partition function for
different quark masses and topological charge $\nu$.

In terms of the eigenvalues the partition function (\ref{zrandom}) is given by
\be
Z_{N_f, \nu}^{\beta =2}(m_1, \cdots, m_{N_f}) = 
\int d\lambda \Delta^2(\lambda_k^2)
\prod_k \lambda_k^{2(N_f+\nu)+1} \prod_f m_f^\nu 
\prod_{f,k}(\lambda_k^2 + m^2_f)
e^{-\frac{\Sigma^2 N}2 \sum_k\lambda_k^2},\nonumber\\
\label{zeig}
\ee
where the Vandermonde determinant is defined by
\be
\Delta(\lambda_k^2) = \prod_{k,l} (\lambda_k^2-\lambda_l^2).
\label{vandermonde}
\ee
By inspection we have
\be
\frac{Z_{N_f,\nu}(m_1, \cdots, m_{N_f})}{ \prod_f m_f^\nu}\sim 
Z_{N_f+\nu,0}(m_1, \cdots, m_{N_f},0,\cdots, 0),
\ee
where the argument of the last factor has $\nu$ zeros.

As an example, the simplest nontrivial identity of this type is given by
\be
Z_{1,1}(m) \sim m Z_{2,0}(m,0).
\ee
Let us  prove this identity without relying on Random 
Matrix Theory. 
According to the definition (\ref{zm}) we have
\be 
Z_{1,1}(m) \sim \int d\theta e^{i\theta} e^{mV\Sigma \cos \theta},
\ee
and 
\be
Z_{2,0}(m,0) \sim \int_{U\in U(2)} dU e^{V \Sigma {\rm Re \,Tr} (MU)},
\ee
where $M$ is a diagonal matrix with diagonal elements $m$ and 0.
The integral over $U$ can be performed by diagonalizing $U$ according
to $U = U_1 e^{i\phi_k} U_1^{-1}$, and choosing $U_1$ and $\phi_k$ as new
integration variables. The Jacobian of this transformation is
\be
J \sim \Delta^2(e^{i\phi_k}).
\ee
The integral over $U_1$ can be performed using the Itzykson-Zuber formula.
This results in
\be
Z_{2,0} \sim \int d\phi_1 d\phi_2 \frac{|e^{-i\phi_1} -e^{-i\phi_2}|^2}
{m(\cos \phi_1 - \cos \phi_2)}
(e^{mV\Sigma \cos \phi_1} - e^{mV\Sigma \cos \phi_2})
\ee
Both terms in the last factor result in the same contribution to the 
integral. Let us consider
only the first term $\sim \exp(mV\Sigma \cos \phi_1)$. Then the integral
over $\phi_2$ has to be defined as a principal value integral. If we use the 
identity
\be
\frac{|e^{-i\phi_1} -e^{-i\phi_2}|^2}
{(\cos \phi_1 - \cos \phi_2)} = 2\left ( \cos \phi_1
-\frac{\sin \phi_1}{\tan ((\phi_1+\phi_2)/2)}\right )
\ee
the $\phi_2$-integral of the term proportional to $\sin\phi_1$ gives zero 
because of the principal value prescription. The term proportional to $\cos 
\phi_1$ trivially results in
$Z_{1,1}$. We leave it as an exercise
to the reader to generalize this proof to arbitrary $N_f$ and $\nu$.

The group integrals in finite volume partition function (\ref{zm})
were evaluated by Leutwyler and Smilga\refnote{\cite{LS}} for $equal$ 
quark masses. An expression for $different$ quark masses was obtained
by Sener{\it  et al.}\refnote{\cite{Sener3}}. The expression could only
be proved for $\nu = 0$. The above duality can be used to relate
a partition function at arbitrary $\nu$ to a partition function at $\nu =0$.
This completes the proof of the conjectured expression for arbitrary 
topological charge.

\section{UNIVERSALITY IN CHIRAL RANDOM MATRIX THEORY}

In the chiral ensembles, two types of universality studies can be performed.
First, the 
universality of correlations in the bulk of the spectrum. 
As discussed above, they are given by the invariant random matrix ensembles.
Second, the universality of the 
microscopic spectral density and the
eigenvalue correlations near zero virtuality. The aim of such studies
is to show that 
microscopic correlations are stable against deformations of the chiral 
ensemble away from the Gaussian probability distribution. 
Recently, a number of universality studies
on microscopic correlations have appeared. They will be
reviewed in this section.

In a first class of universality studies one considers probability
distributions that maintain unitary invariance, i.e.
\be
P(W) \sim \exp(-N\beta \frac {\Sigma^2 }4 \sum_{k=1}^\infty a_k {\rm Tr} 
(W^\dagger W)^k).
\ee
The first study of this kind was performed by Br\'ezin, Hikami and 
Zee\refnote{\cite{brezin-hikami-zee}}. 
They considered
a potential with only $a_1$ and $a_2$ different from zero and showed that
the microscopic spectral density is independent of $a_2$. A general proof 
valid for arbitrary potential was given by 
Nishigaki\refnote{\cite{Nishigaki-uni,Nishigaki-zak}}. 
The essence of the proof
is a remarkable generalization of the identity for
the Laguerre polynomials,
\be
\lim_{n \rightarrow \infty}  L_n(\frac x n) =
 J_0(2 \sqrt x) \ ,
\label{asym}
\ee
to orthogonal polynomials determined by an arbitrary potential $V$. 
This relation was proved by deriving a differential equation
from the continuum limit of the recursion relation for orthogonal polynomials.
In a consecutive work, Akemann, Damgaard, Magnea and 
Nishigaki\refnote{\cite{Damgaard}}  extended this
proof to all microscopic correlation functions.

In a second class of universality studies one considers 
deformations of the Gaussian random matrix
ensemble that violate unitary invariance. In particular, one has 
considered the case where the 
matrix $W$ in (\ref{diracop}) is replaced by
\be 
W\rightarrow W+ A,
\label{weps}
\ee
where, because of the unitary invariance,  the matrix $A$ 
can always be chosen diagonal.
The simplest case with 
$A=\pi T$ times the identity was
considered by Sener {\it et al.}\refnote{\cite{Sener1}}. 
This model provides a schematic model
of the chiral phase transition. For large matrices, the average 
resolvent defined by
\be
G(z) = {\rm Tr} \langle  \frac 1{z+i\epsilon -{\cal D}}\rangle
\ee
obeys the cubic equation\refnote{\cite{JV,Stephanov1,Sener1}} (the parameter
$\Sigma = 1$ in (\ref{zrandom}))
\be
G^3 -2zG^2 +G(z^2-\pi^2T^2 + 1) - z = 0.
\ee
This equation was first obtained using a diagrammatic 
method\refnote{\cite{JV,Sener1}}. Later, this derivation was 
rewritten\refnote{\cite{nowakblue}} in terms of the so called blues 
function\refnote{\cite{zeeblue}}. 
The average spectral density given by
\be
\rho(\lambda) = -\frac 1\pi {\rm Im} G(z=\lambda)
\ee 
is a semicircle at $\pi T=0$ and splits into two arcs at $\pi T = 1$. For 
the spectral density at zero one obtains 
$\rho(0) = \sqrt{1-\pi^2 T^2}$, and therefore chiral
symmetry is broken for $\pi T < 1$, and is restored above this temperature.
In spite of this drastic change in average spectral density, it could
be shown\refnote{\cite{Sener1}} with the help of a supersymmetric formulation
of Random Matrix Theory that the microscopic
spectral density does not depend on $T$.

The super-symmetric method in the first paper by 
Sener $et$ $al.$\refnote{\cite{Sener1}} is not
easily generalizable to higher order correlation functions. A natural
way to proceed is to employ the super-symmetric method introduced by 
Guhr\refnote{\cite{super-thomas}}. In the case of $\beta = 2$ this 
method results in an expression for the kernel determining all 
correlation functions. This approach was followed in two papers, one by
Guhr and Wettig\refnote{\cite{GWu}} and one by 
Sener {\it et al.}\refnote{\cite{Seneru}}. 
The latter authors studied 
microscopic correlation functions for $A$ in (\ref{weps}) 
proportional to the identity, whereas Guhr and Wettig  considered
an arbitrary diagonal matrix $A$. It was shown that independent
of the matrix $A$, the correlations are given by the 
Bessel kernel\refnote{\cite{Tracy}}. Of course, a necessary condition on
the matrix $A$ is that chiral symmetry is broken.
Guhr and Wettig also showed that correlations in the bulk of the 
spectrum are insensitive to $A$. 

The main ingredient of the proof was a supersymmetric 
generalization\refnote{\cite{Guhr-Wettig}} of
the Itzykson-Zuber-Harish-Chandra type 
integral\refnote{\cite{Guhr-Wettig,Sener3,Berezin}}
\be
\int d\mu(U,V) \exp({\rm Re\, Tr}U^\dagger S V R) = C
\frac {\det_{k,l}(I_\nu(r_k s_l))}{ \prod_{k=1}^{N_2}(r_ks_k)^\nu
\Delta( S^2)\Delta( R^2)}.
\ee
Here, $U \in U(N_1)$, $V \in U(N_2)$, $\nu = N_1-N_2 \ge 0$. The positive
square roots of the eigenvalues of $S^\dagger S$ and $R R^\dagger$ are 
denoted by, $s_k$ and $r_k$, respectively. The integral is over the invariant
measure. The constant $C$ in the r.h.s. can be evaluated, and the $I_\nu$
are modified Bessel functions.

The deformation $W\rightarrow W+A$ with the probability distribution 
for $W$ given by an arbitrary 
invariant potential has not yet been considered. 
We have no doubt that universality proofs along the lines of
methods developed by P. Zinn-Justin\refnote{\cite{Pzinn}}
can be given.  

We wish to emphasize that all universality studies for the chiral ensembles
have been performed for
$\beta =2$. The reason is that $\beta =1$ and $\beta=4$ are mathematically 
much more involved. It certainly would be of interest to extend the
above results to these cases as well.

In addition to the above analytical studies the 
universality of the microscopic spectral density also follows from
numerical studies of models with the symmetries of the QCD partition function.
In particular, we mention strong
support in favor universality 
from a different branch of physics, namely from
the theory of universal conductance fluctuations. In that context, the
microscopic spectral density of the eigenvalues of the transmission matrix
was calculated for the Hofstadter\refnote{\cite{HOFSTADTER}}
model, and, to a high degree of accuracy, it agrees with
the random matrix prediction\refnote{\cite{SLEVIN-NAGAO}}.
Other studies deal with a class random matrix models with matrix elements with
a diverging variance.
Also in this case the microscopic spectral density is given
by the universal expressions\refnote{\cite{Kelner}} 
(\ref{micro1}) and (\ref{micro2}). 

The conclusion that emerges from all numerical and analytical 
work on modified chiral random matrix models
is that the microscopic
spectral density  and the correlations near zero virtuality 
exhibit a strong universality that is comparable to the stability 
of microscopic correlations in the bulk of the spectrum.

Of course, QCD is much richer than chiral Random Matrix Theory. 
One question that
should be asked is at what scale (in virtuality) QCD spectral correlations
deviate from RMT. 
This question has been studied by means of instanton liquid
simulations. Indeed, at macroscopic scales, it was 
found that the number variance
shows a linear dependence instead of the logarithmic dependence 
observed at microscopic scales\refnote{\cite{Osborninst}}. 
More work is needed to determine the point where the crossover between
these two regimes takes place.

\section{LATTICE QCD RESULTS}

Recently, the Dirac spectrum in lattice QCD received a good deal of attention. 
In particular, the connection between the topology of
field configurations and the spectrum of the Wilson Dirac operator 
has been studied in detail\refnote{\cite{Ivanov,Bardeen,Sailer}}.
Other studies are related to the connection between the Wilson Dirac
spectrum and the localization properties of the 
eigenfunctions\refnote{\cite{Janssen}}. 

In this section we will focus ourselves on the spectral correlations of the
lattice QCD Dirac operator. Both 
correlations in the bulk of the spectrum and the microscopic spectral
density will be studied. Consistent with 
universality arguments presented above, we
find that spectral correlations are in complete agreement with
chiral Random Matrix Theory.

\subsection{Correlations in the Bulk of the Spectrum}
 
Recently, Kalkreuter\refnote{\cite{Kalkreuter}}
calculated $all$ eigenvalues of the $N_c = 2$ lattice Dirac
operator both for Kogut-Susskind (KS) fermions and Wilson fermions
for lattices 
as large as $12^4$. For the
Kogut-Susskind Dirac operator, $D^{KS}$, we use the convention
that it is
anti-Hermitean. Because of the Wilson-term, the Wilson Dirac operator, $D^W$,
is neither Hermitean nor anti-Hermitean. Its 
Hermiticity relation is given by
${D^W}^\dagger = \gamma_5 D^W \gamma_5$.
 Therefore, the operator
$\gamma_5 D^W$ is Hermitean. However, it does not anti-commute with $\gamma_5$,
and its eigenvalues do not occur in pairs $\pm \lambda_k$. 

\begin{center}
\begin{figure}[!ht]
\centering\includegraphics[width=115mm,angle=0]
{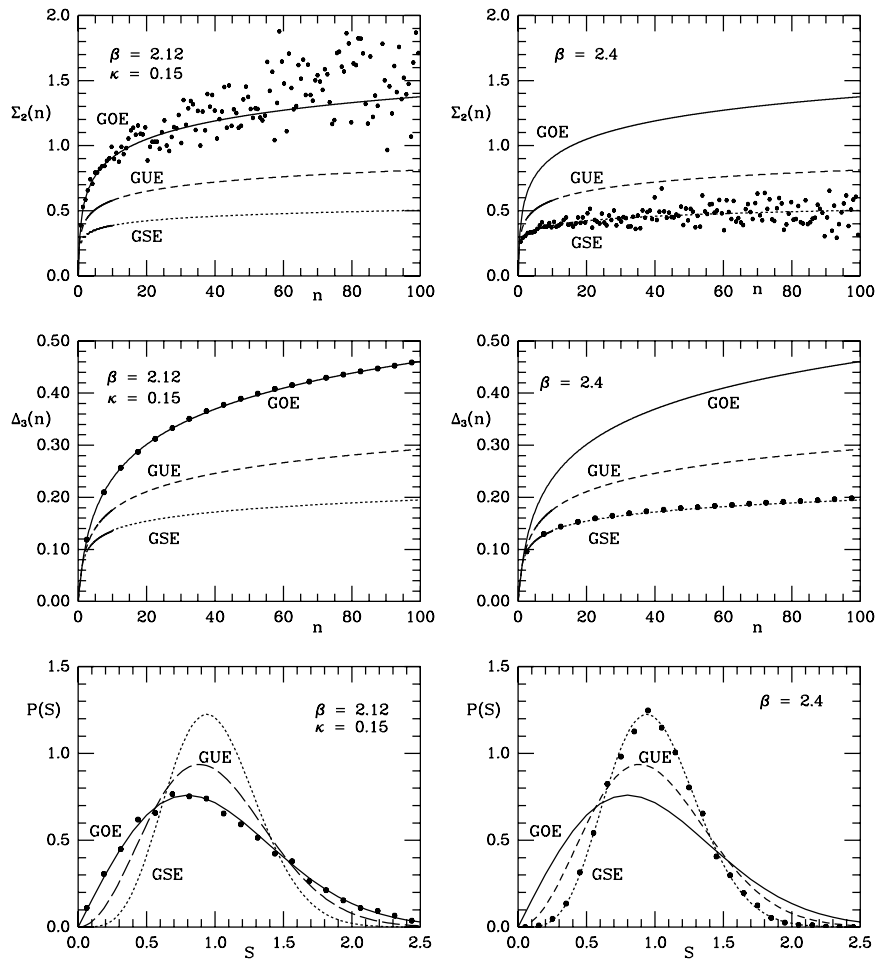}
\caption{
Spectral correlations of Dirac eigenvalues for Wilson fermions
(left) and KS-fermions (right). Results are shown for the number variance,
$\Sigma_2(n)$, the $\Delta_3-$statistic and the nearest neighbor spacing
distribution, $P(S)$. The full, dashed and dotted curves represent the
analytical result for the GOE, GUE and GSE, respectively.}
\label{fig2}
\end{figure}
\end{center}
\noindent
In the the case of $SU(2)$,
the anti-unitary symmetry of the Kogut-Susskind 
and the Wilson Dirac operator is given by\refnote{\cite{Teper,HV}},
\be
[D^{KS}, \tau_2 K] = 0, \quad {\rm and} \quad 
[\gamma_5 D^W, \gamma_5 CK\tau_2] = 0.
\ee
Because
\be
(\tau_2 K)^2 = -1 ,\quad {\rm and} \quad (\gamma_5 CK\tau_2)^2 =1,
\ee
the matrix elements of the KS Dirac operator can be arranged into 
real quaternions, whereas the Wilson Dirac operator can be expressed
into real matrix elements.
Therefore, we expect that eigenvalue
correlations in the bulk of the spectrum 
are described by the GSE and the GOE, respectively\refnote{\cite{HV}}.
The microscopic correlations for KS fermions are described by the chGSE. However,
the microscopic correlations for Wilson fermions are not described by
the chGOE but rather by the GOE.  
Because of the anti-unitary symmetry, the 
eigenvalues of the KS Dirac operator are subject to the Kramers degeneracy,
i.e. they are double degenerate.

In both cases, the Dirac matrix is tri-diagonalized by
Cullum's and Willoughby's Lanczos procedure\refnote{\cite{cullum}} 
and diagonalized with a standard
QL algorithm. This improved algorithm makes it possible to obtain $all$
eigenvalues. This allows us to test the accuracy of the eigenvalues
by means of sum-rules for the sum of the squares of the eigenvalues
of the lattice Dirac operator. Typically, the numerical error in the
sum rule is of order $10^{-8}$. 

\begin{center}
\begin{figure}[!ht]
\centering\includegraphics[width=60mm,angle=0]
{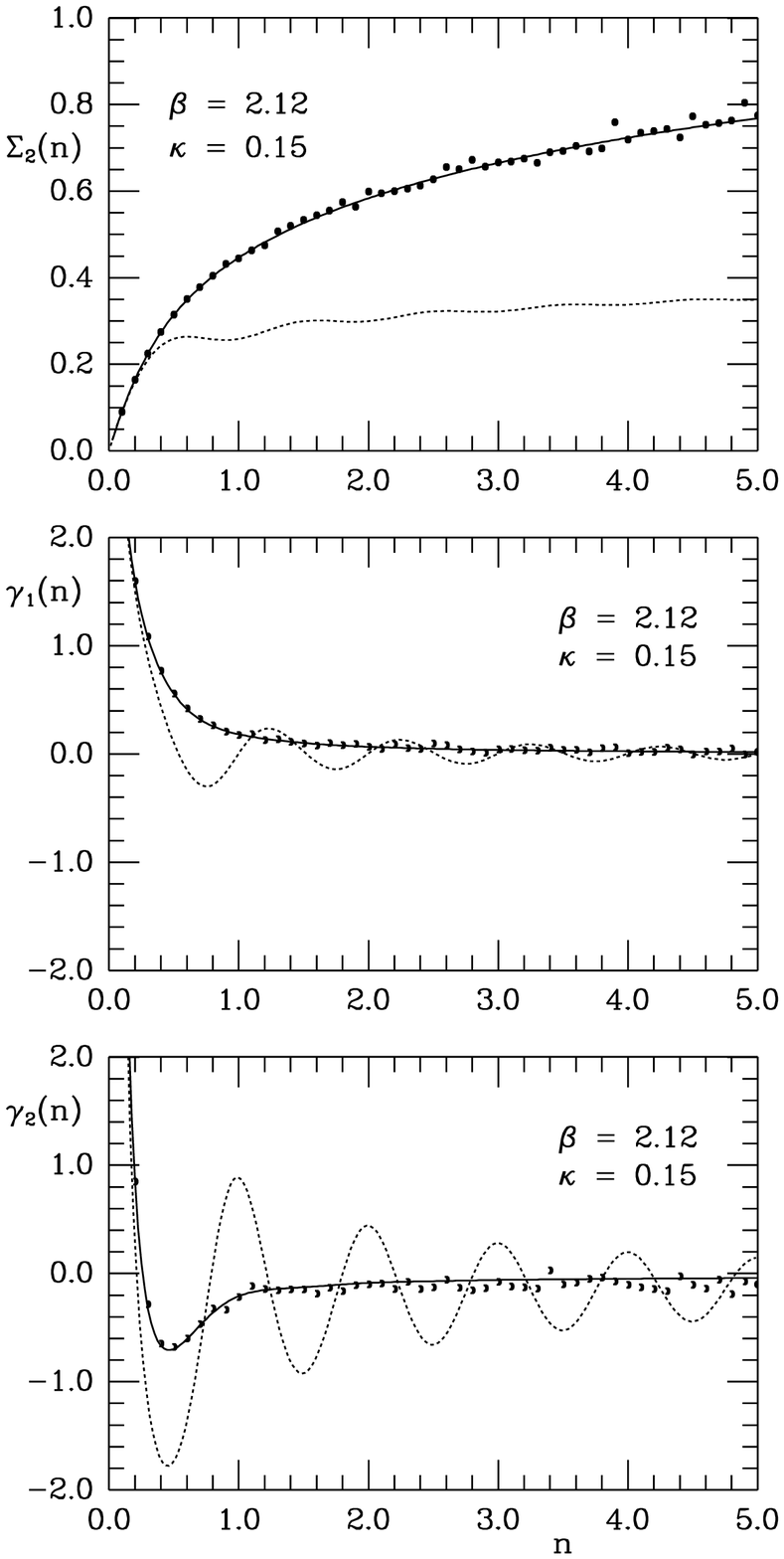}
\includegraphics[width=60mm,angle=0]
{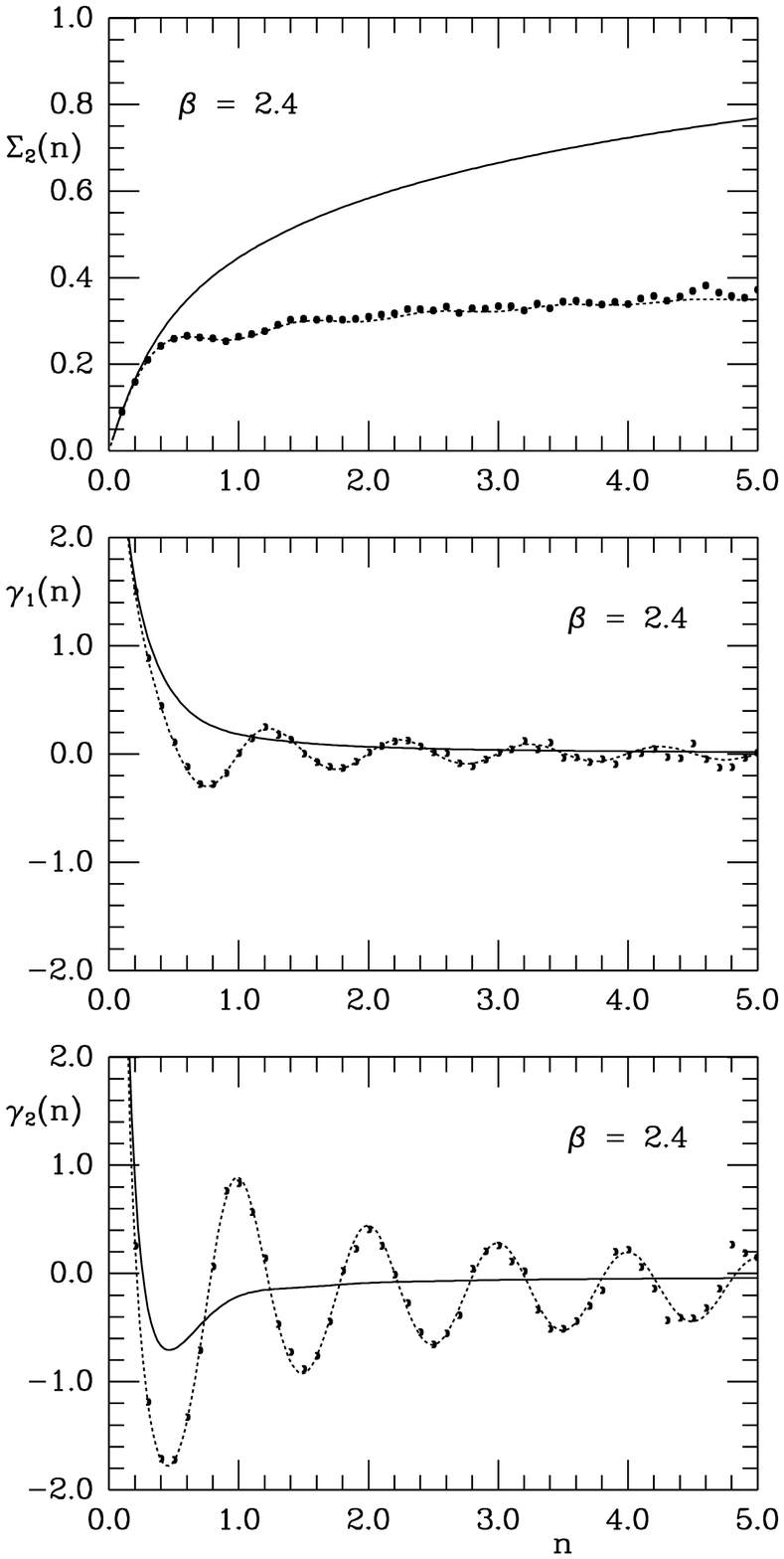}
\caption{The number variance, $\Sigma_2(n)$ and the first two cumulants,
$\gamma_1(n)$ and $\gamma_2(n)$ as a function of $n$ for  eigenvalues
of the Wilson Dirac operator (left) and the Kogut-Susskind Dirac operator 
(right). The full and dotted
curves represent the analytical result for the GOE and the GSE, respectively.}
\label{fig3}
\end{figure}
\end{center}
\noindent

As an example, in Fig. 1 we show a histogram 
of the overall Dirac spectrum for KS fermions at 
$\beta = 2.4$. Results for the spectral correlations are shown in Figs. 2 and 
3. The results for KS fermions are for 4 dynamical flavors
with $ma = 0.05$ on a $12^4$ lattice. The results for Wilson fermions were
obtained for two dynamical flavors on a $8^3\times 12$ lattice. For
the values of $\beta$ and $\kappa$ we refer to the labels of the figure.
For $\beta > 2.4$, with our lattice parameters for KS fermions,
the Dirac spectrum near zero virtuality develops a gap. Of course, this is 
an expected feature of the weak coupling domain. 
For small enough value of $\kappa$
the Wilson Dirac spectrum shows a gap at $\lambda = 0$ as well. In the scaling 
domain the value of $\kappa$ is just above the critical value of $\kappa$.
A qualitative description of the Wilson Dirac spectrum can be obtained
with a random matrix model with the structure of the Wilson Dirac 
operator\refnote{\cite{Jurkiewicz}}.

The eigenvalue spectrum is unfolded by fitting a second order polynomial to the
integrated spectral density of a stretch of 500-1000 eigenvalues.
The results for  $\Sigma_2(n)$,
$\Delta_3(n)$ and $P(S)$ in Fig. 2 show an impressive agreement with the 
RMT predictions. The fluctuations in $\Sigma_2(n)$ are as expected from RMT.
The advantage of $\Delta_3$-statistic is well illustrated by this figure. 
We also investigated\refnote{\cite{HKV}} the $n$ dependence of the
first two cumulants of the number of levels in a stretch of length $n$. Results
presented in Fig. 4 show a perfect agreement with RMT.

Spectra for different values of $\beta$ have been analyzed as well. 
It is probably no
surprise that random matrix correlations are found at stronger couplings.
What is surprising, however, is that even in the 
weak-coupling domain ($\beta =2.8$) the
eigenvalue correlations are in complete agreement with Random
Matrix Theory. Finally, we have studied the stationarity of the ensemble
by analyzing level sequences of about 200 eigenvalues (with relatively
low statistics). No deviations from random matrix correlations were observed
all over the spectrum, including the region near  $\lambda = 0$. 
This justifies the spectral averaging which results in
the good statistics in Figs. 2 and 3.

In the case of three or more colors with fundamental fermions, both the 
Wilson and Kogut-Susskind Dirac operator do not possess any anti-unitary 
symmetries. Therefore, our conjecture is that in this case
the spectral correlations in the bulk of the spectrum of 
both types of fermions can be described by
the GUE. In the case of two fundamental colors the continuum theory 
and Wilson fermions are in the same universality class.
It is an interesting question of how spectral correlations of KS fermions
evolve in the approach to the continuum limit. Certainly, the 
Kramers degeneracy of the eigenvalues remains. However, since Kogut-Susskind
fermions represent 4 degenerate flavors in the continuum limit, 
the Dirac eigenvalues should obtain an additional two-fold degeneracy.
We are looking forward to more work in this direction.

\subsection{The Microscopic Spectral Density}

The advantage of studying spectral correlations in the bulk of the
spectrum is that one can perform spectral averages instead of ensemble
averages requiring only a relatively small number of equilibrated
configurations. This so called spectral ergodicity cannot
be exploited in the study of the microscopic spectral density.
In order to gather sufficient statistics for the microscopic
spectral density of the lattice Dirac operator
a large number of independent configurations is 
needed. One way to proceed is to generate instanton-liquid configurations
which can be obtained much more cheaply than lattice QCD configurations.
Results of such analysis\refnote{\cite{Vinst}} show that for $N_c=2$
with fundamental fermions the microscopic spectral density is given
by the chGOE. For $N_c =3$ it is given by the chGUE. One could argue that
instanton-liquid configurations can be viewed as 
smoothened lattice QCD configurations. Roughening such configurations
will only improve the agreement with Random Matrix Theory. 

Of course, the ultimate goal is to test the conjecture of microscopic
universality for realistic lattice QCD configurations. In order to obtain a
very large number of independent gauge field configurations 
one is necessarily restricted to relatively
small lattices. The first study in this direction was reported 
recently\refnote{\cite{berbenni,tilohirsch}}. In this work, the 
quenched $SU(2)$ Kogut-Susskind
Dirac operator is diagonalized for lattices with linear dimension of
4, 6, 8 and 10, and a total number of configurations of 9978, 9953, 3896 and
1416, respectively. The results were compared with predictions from the 
chGSE.

\begin{center}
\begin{figure}[!ht]
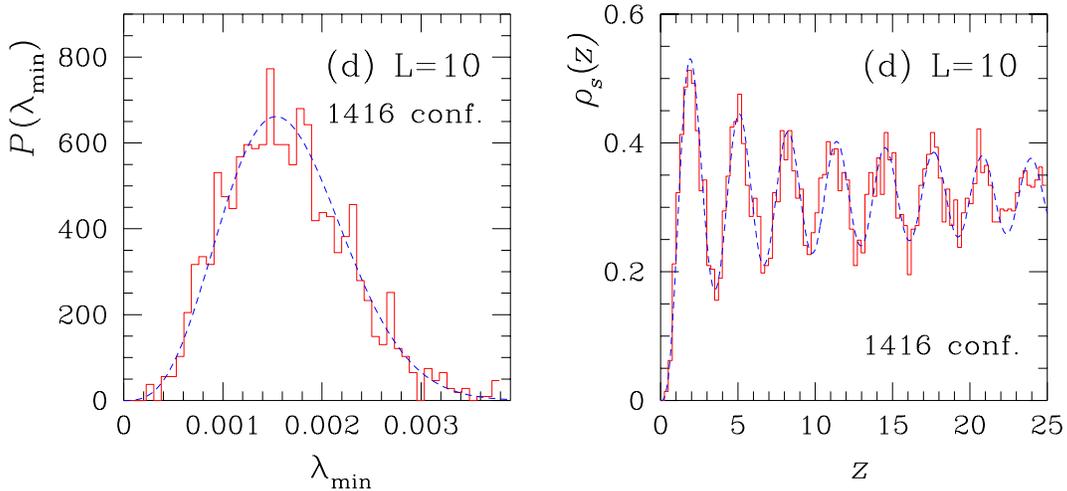

\vspace*{-1cm}
\centering\includegraphics[width=70mm,angle=0]
{tilo10small.ps}
\includegraphics[width=70mm,angle=0]
{tilo10micro.ps}
\vspace*{-1.5cm}
\caption{The distribution of the smallest eigenvalue (left) and the
microscopic spectral density (right) of the Kogut-Susskind Dirac operator
for two colors and $\beta = 2.0$. Lattice results are represented by
the histogram, and the analytical results for the chGSE are given by the
dashed curves.}
\label{fig4}
\end{figure}
\end{center}
\noindent

\noindent
We only show results for the largest lattice. For more detailed results,
including results for the two-point correlation function,
we refer to the original work. In Fig. 4  we show the distribution of 
the smallest eigenvalue (left) and the microscopic spectral density (right).
The lattice results are given by the full line. The dashed curve represents
the random matrix results. The distribution of the smallest eigenvalue was
derived by Forrester\refnote{\cite{Forrester}} and is given by
\be
P(\lambda_{\rm min}) =\alpha \sqrt{\frac \pi{2}} (\alpha\lambda_{\rm min})^{3/2}
I_{3/2}(\alpha\lambda_{\rm min}) e^{-\frac 12 (\alpha\lambda_{\rm min})^2},
\ee
where $\alpha =V \Sigma$.
The random matrix result for the microscopic spectral density is given
in eq. (\ref{micro4}). We emphasize that the theoretical curves have been
obtained without any fitting of parameters. The input parameter, the
chiral condensate, is derived from the same lattice calculations. 
The above simulations were performed at a relatively strong coupling of
$\beta = 2$. Recently, the same analysis\refnote{\cite{tilopr}} 
was performed for  $\beta = 2.2$
and for $\beta =2.5$ on a $16^4$ lattice. In both cases
agreement with the random matrix predictions was found\refnote{\cite{tilopr}}.

An alternative way to probe the Dirac spectrum is via
the valence quark mass dependence of the chiral 
condensate\refnote{\cite{Christ}} defined as
\be
\Sigma(m) = \frac 1N \int d\lambda \langle\rho(\lambda)\rangle
\frac{2m}{\lambda^2 +m^2}.
\label{sigmam}
\ee
The average spectral density is obtained for a fixed sea quark mass.
For masses well beyond the smallest eigenvalue, $\Sigma(m)$ shows a plateau
approaching the value of the chiral condensate $\Sigma$. 
In the mesoscopic range (\ref{range}), we can introduce $u=\lambda m N$ and
$x= mN\Sigma$ as new variables. Then 
the microscopic spectral density enters in $\Sigma(m)$.
For three fundamental colors the microscopic spectral density
for $\beta =2$ (eq. (\ref{micro2})) applies and the integral over $\lambda$
in (\ref{sigmam}) can be performed analytically. The result is given 
by\refnote{\cite{VPLB}},
\be
\frac {\Sigma(x)}{\Sigma} = x(I_{a}(x)K_{a}(x)
+I_{a+1}(x)K_{a-1}(x)),
\label{val}
\ee
where $a = N_f+|\nu|$.
In Fig. 2 we plot this ratio as a function of $x$ (the 'volume' $V$ is equal
to the total number of Dirac eigenvalues) for
lattice data of two dynamical flavors with  mass $ma = 0.01$ and $N_c= 3$ on a
$16^3 \times 4$ lattice.  We observe
that the lattice data for different values of $\beta$ fall on a single curve.
Moreover, in the mesoscopic range 
this curve coincides with the random matrix prediction for $N_f = \nu = 0$.
\begin{center}
\begin{figure}[!ht]
\centering\includegraphics[width=75mm,angle=270]
{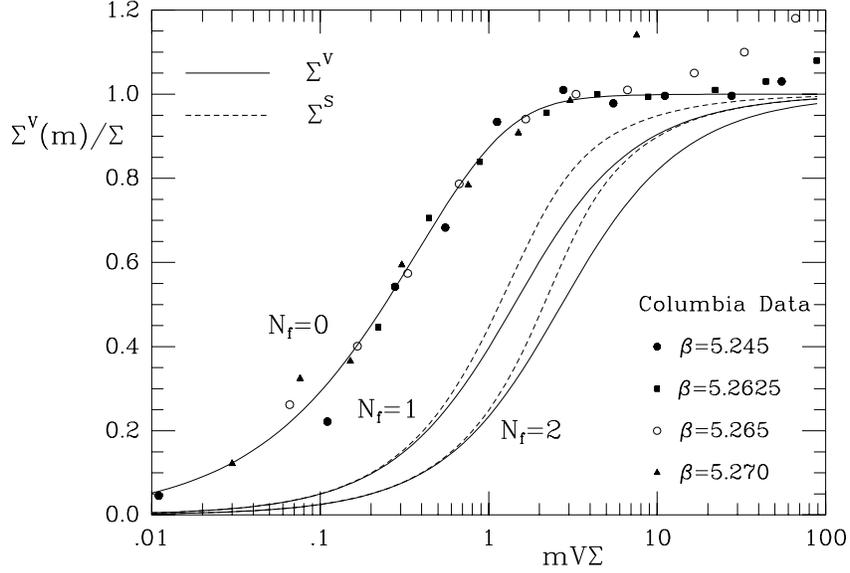}
\caption{The valence quark mass dependence of the chiral condensate
$\Sigma^V(m)$ plotted as $\Sigma^V(m)/\Sigma$ versus $m V\Sigma$. The dots and
squares represent lattice results by the Columbia group\refnote{\cite{Christ}}
 for values of $\beta$ as indicated in the label of the figure.}
\label{fig5}
\end{figure}
\end{center}
Apparently, the zero modes are completely mixed with the much larger number of
nonzero modes. For eigenvalues much smaller than the sea quark mass, one 
expects quenched ($N_f = 0$) eigenvalue correlations. 
In the same figure the dashed curves represent results for the quark mass 
dependence of the
chiral condensate (i.e. the mass dependence for equal valence and sea quark 
masses). 
In the sector of zero topological charge 
one finds\refnote{\cite{jolio,GL,LS}}
\be
\frac{\Sigma^S(u)}{\Sigma} = \frac{I_1(u)}{I_0(u)}\quad {\rm for}\quad
N_f = 1,
\label{sigmam1f}
\ee
and
\be
\frac{\Sigma^S(u)}{\Sigma} = 
\frac{I_1^2(u)}{u(I_0^2(u)-I_1^2(u))}\quad {\rm for}\quad N_f = 2.
\ee
We observe that both expressions do not fit the data. Also notice that, 
according to G\"ockeler {\it  et al.}\refnote{\cite{Gockeler}},
eq. (\ref{sigmam1f}) describes the valence mass dependence of the
chiral condensate for non-compact QED with quenched Kogut-Susskind fermions. 
However, we were not able to derive their result (no derivation is given in 
the paper).

\section{CHIRAL RANDOM MATRIX THEORY AT $\mu \ne  0$}
At nonzero temperature $T$ 
and chemical potential $\mu$ a $schematic$ random matrix 
matrix model of the QCD partition function is obtained by replacing
the Dirac operator in (\ref{zrandom}) 
by\refnote{\cite{JV,Tilo,Stephanov1,Stephanov}}
\be
{\cal D} = \left (\begin{array}{cc} 0 & iW +i\Omega_T +\mu \\
iW^\dagger+i\Omega_T + \mu & 0 \end{array} \right ).
\label{diracmattert}
\ee
Here, $\Omega_T= T \otimes_n(2n+1)\pi {\large\bf 1}+\mu$  
are the matrix elements
of $\gamma_0 \partial_0 +\mu \gamma_0$ in a plane wave basis with 
anti-periodic boundary conditions in the time direction.
Below, we will discuss a model with $\Omega_T$ 
absorbed in the random matrix and $\mu \ne 0$. The aim of this
model is to explore the effects of the non-Hermiticity of the 
Dirac operator. For example, 
the random matrix partition function (\ref{zrandom}) 
with the Dirac matrix (\ref{diracmattert}) is well suited for
the study of zeros of this partition function in the complex
mass plane and in the complex chemical potential plane.
For a complete analytical understanding of the location of such zeros
we refer to the work by Halasz {\it et al.}\refnote{\cite{Halaszyl}}. 

The term $\mu\gamma_0$ does not affect the anti-unitary
symmetries of the Dirac operator. This is also the case in lattice QCD
where the color matrices in the forward time direction are replaced
by $U \rightarrow e^\mu U$ and in the backward time direction by
$U^\dagger \rightarrow e^{-\mu} U^\dagger$. For this reason the 
universality classes are the same 
as at zero chemical potential. 

The Dirac
operator that will be discussed in this section is thus given by
\be
{\cal D}(\mu) = \left ( \begin{array}{cc} 0 & iW +\mu \\
iW^\dagger+ \mu & 0 \end{array} \right ),
\label{diracmatter}
\ee
where the matrix elements $W$ are either real ($\beta = 1$), 
complex ($\beta = 2$) or quaternion real ($\beta = 4$). 
For all three values of $\beta$ the eigenvalues of
${\cal D}(\mu)$ are scattered in the complex plane.

Since many standard random matrix methods 
rely on convergence properties based on the
Hermiticity of the random matrix, direct 
application of most methods is not possible. The simplest way out is
the Hermitization\refnote{\cite{Feinberg-Zee}} 
of the problem, i.e we consider
the Hermitean operator
\be
D^H(z,z^*) = \left ( \begin{array}{cc} \kappa & z-{\cal D}(\mu) \\
z^* - {\cal D}^\dagger(\mu) & \kappa \end{array} \right ).
\label{diracherm}
\ee
For example, the generating function in the supersymmetric method of Random
Matrix Theory\refnote{\cite{Efetov,VWZ}} 
is then given by\refnote{\cite{supernonh,Khoruzhenko,fyodorov,Efetovnh}}
\be
Z(J,J^*) = \langle \frac {\det(D^H(z+J,z^*+J^*) }{\det(D^H(z,z^*))} 
\rangle.
\label{generator}
\ee
The determinants can be rewritten as fermionic and bosonic
integrals. Convergence is assured by the Hermiticity and the infinitesimal
increment $\kappa$.
The resolvent follows from the generating function by differentiation
with respect to the source terms
\be
G(z,z^*) = {\rm Tr }\langle \frac 1{z-{\cal D}(\mu)} \rangle= 
\left . \frac {\partial}{\partial J} Z(J,J^*)\right |_{J=J^* = 0}.
\ee
Notice that, after averaging over the random matrix, the partition function
depends in a non-trivial way on both $z$ and $z^*$. 
The spectral density is then given by
\be
\rho(z) = \frac 1\pi \frac {\partial}{\partial z^*} G(z).
\ee
Alternatively, one can use the replica 
trick\refnote{\cite{Girko,Stephanov}} with
generating function given by
\be
\langle {\det}^{N_f} D_H(z,z^*) \rangle.
\label{replica}
\ee 
The idea is to perform the calculation for integer values
of $N_f$ and perform the limit $N_f \rightarrow 0$ at the end of the
calculation. 
Although the replica limit, $N_f \rightarrow 0$, fails in 
general\refnote{\cite{martinrep}}, it is expected to work for
$\det D_H(z,z^*)$ because it is positive definite (or zero). Then the
partition function is a smooth function of $N_f$.
For other techniques addressing nonhermitean matrices we refer to the 
recent papers by Feinberg and Zee\refnote{\cite{Feinberg-Zee}} and Nowak and
co-workers\refnote{\cite{Nowakrecent}}.
One recent method that does not rely on the Hermiticity of the random
matrices is the method of complex orthogonal 
polynomials\refnote{\cite{fyodorovpoly}}. This method
was used by Fyodorov {\it et al.}\refnote{\cite{fyodorovpoly}} to
calculate the number variance and the nearest
neighbor spacing distribution in the regime of weakly nonhermitean matrices.
As surprising new result, they found an $S^{5/2}$ repulsion law.
 
In the physically relevant case of QCD with three colors,  
the fermion determinant
is complex for nonzero chemical potential. Its 
phase prevents the convergence of fully unquenched Monte-Carlo
simulations (see Kogut {\it et al.}\refnote{\cite{Klepfish}} for the latest
progress in this direction). However, it is possible to perform quenched
simulations. In such calculations it was found that the critical
chemical potential $\mu_c \sim \sqrt m$, instead of a third of the nucleon
mass\refnote{\cite{everybody}}. This phenomenon was explained
analytically by Stephanov\refnote{\cite{Stephanov}} with the help of the
above random matrix model. He could show that for small $\mu$
the eigenvalues are distributed along the imaginary axis in a band of width
$\sim \mu^2$ leading to a critical chemical potential of $\mu_c \sim m^2$.
As has been argued above, the quenched limit 
is necessarily obtained from a partition function in which the 
fermion determinant appears as
\be
\lim_{N_f \rightarrow 0} | \det {\cal D}(\mu)|^{N_f},
\ee
instead of the same expression without the absolute value signs. The
partition function with the absolute value of the determinant can be
interpreted as a partition function of an equal number of fermions and
conjugate fermions. The critical value of the chemical potential, equal
to half the pion mass, is
due to Goldstone bosons with net a baryon number
consisting of a quark and conjugate quark.
The reason that the quenched limit does not correspond to the standard
QCD partition function is closely related to the failure of the replica
trick in the case of a determinant with a nontrivial phase.

Both for $\beta= 1$ and $\beta = 4$ the fermion determinant, 
$\det ({\cal D}(\mu) +m)$, is real. This
is obvious for $\beta = 1$. For $\beta = 4$ the reality follows from the 
identity $q^* = \sigma_2 q \sigma_2$ for a quaternion real element $q$, and
the invariance of a determinant under transposition. 
We thus conclude that quenching works for an even number of flavors.
Consequently, chiral symmetry will be  restored for arbitrarily
small nonzero $\mu$, whereas a condensate of
a quark and a conjugate quark develops. Indeed,
this phenomenon has been observed
in the strong coupling limit of lattice QCD
with two colors\refnote{\cite{Elbio}}.

\begin{figure}[ht]
\centering\includegraphics[width=90mm,angle=-90]
{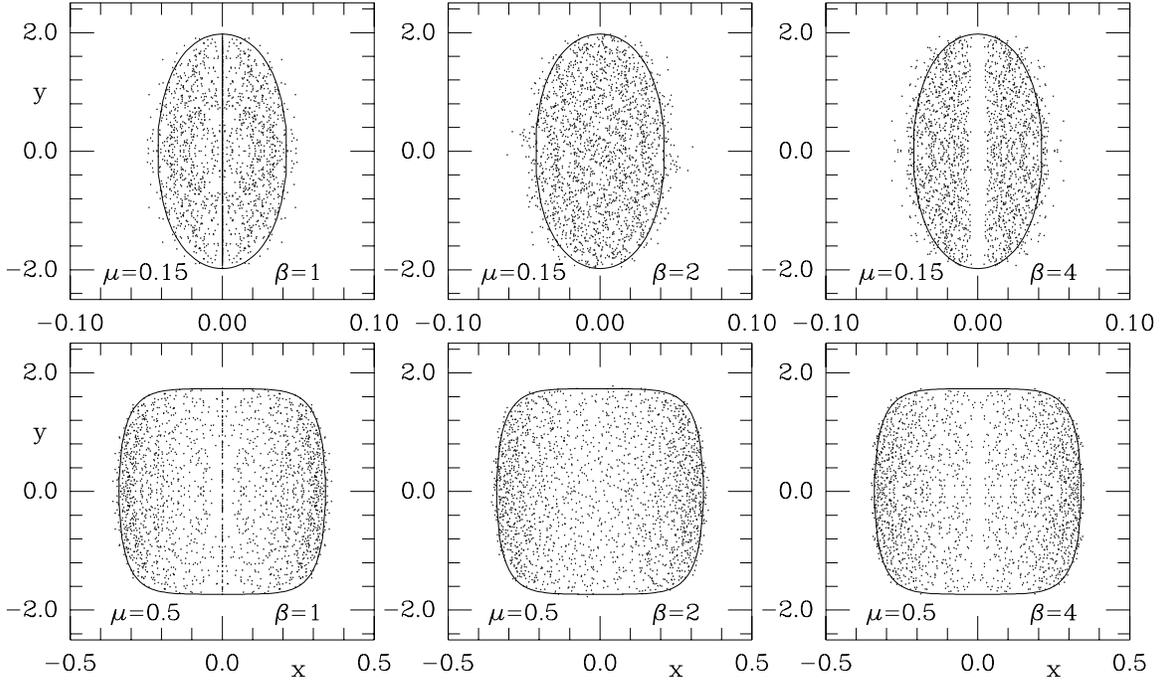}
\caption{
Scatter plot of the real ($x$), and the imaginary
parts ( $y$) of the eigenvalues of the random matrix Dirac operator.
The values of $\beta$ and $\mu$ are given in the labels of the figure.
The full curve shows the analytical result for the boundary.}
\label{fig6}
\end{figure}

In the quenched approximation, the spectral properties
of the random matrix ensemble (\ref{diracmatter})
can be easily studied numerically by simply diagonalizing a set of 
matrices with probability distribution (\ref{zrandom}). 
In Fig. 6 we show numerical results\refnote{\cite{Osborn}} 
for the eigenvalues of a few
$100\times  100$ matrices for $\mu = 0.15$ and $\mu = 0.5$.
The dots represent the eigenvalues in the complex plane.
The full line is the analytical result\refnote{\cite{Stephanov}}
for the boundary of the eigenvalues which is given by
the algebraic curve 
\be
(y^2 +4\mu^2)(\mu^2-x^2)^2 +x^2 = 4\mu^2(\mu^2-x^2).
\label{algebraic}
\ee
This result was obtained by Stephanov for $\beta=2$ 
with the help of the fermionic replica trick. This amounts to
rewriting the determinants in (\ref{replica}) as integrals over 
Grassmann variables. Since Grassmann integrals are always convergent
the infinitesimal increment $\kappa$ can be put equal to zero.
This method can be extended\refnote{\cite{Osborn}} 
to $\beta = 1$ and $\beta =4$. Although the effective partition function
is much more complicated, it can be shown without too much effort that the
solutions of the saddle point equations are the same 
if the variance of the probability distribution is scaled as $1/\beta$.
In particular, the boundary of the domain of eigenvalues 
is the same in each of the three cases. 
However, as one observes from Fig. 6,  for
$\beta =1$ and $\beta = 4$ the spectral density
deviates significantly from the saddle-point result.
For $\beta = 1$
we find an accumulation of eigenvalues on the imaginary axis, whereas for
$\beta = 4$ we find a depletion of eigenvalues in this domain.
This depletion can be understood as follows. For $\mu = 0$ all eigenvalues
are doubly degenerate. This degeneracy is broken at $\mu\ne 0$ which produces
the observed repulsion of the eigenvalues.

The number of purely imaginary eigenvalues for $\beta = 1$
appears to scale as $\sqrt N$. This
explains that this effect is not visible in a leading order saddle
point analysis. From a perturbative analysis
of (\ref{replica}) one obtains a power series in $1/N$.
Clearly, the $\sqrt N$ dependence
requires a truly nonperturbative analysis
of the partition function (\ref{zrandom}) with 
the Dirac operator (\ref{diracmatter}). 
Such a $\sqrt N$  scaling behavior is typical
for the regime of weak non-hermiticity first identified by Fyodorov
{\it et al.}\refnote{\cite{fyodorov}}. 
Using the supersymmetric method for the generating 
function (\ref{generator}) the $\sqrt N$ dependence was obtained
analytically by Efetov\refnote{\cite{Efetovnh}}.

A similar cut below a
cloud of eigenvalues was found in instanton liquid 
simulations\refnote{\cite{Thomas}}  for $N_c =2$ at $\mu \ne 0$ and
in a random matrix model of arbitrary real 
matrices\refnote{\cite{Khoruzhenko}}. The depletion of the eigenvalues
along the imaginary axis was observed earlier in lattice QCD simulations
with staggered fermions\refnote{\cite{baillie}}. 
Obviously, more work has to be done in order to 
arrive at a complete characterization of the 
universal features\refnote{\cite{fyodorovpoly}} in 
the spectrum of nonhermitean matrices.

\section{CONCLUSIONS}
We have argued that there is an intimate relation between correlations of
Dirac eigenvalues and the breaking of chiral symmetry. In the chiral limit,
the fermion determinant suppresses gauge field configurations with 
small Dirac eigenvalues. Correlations counteract this suppression, and
are a necessary ingredient of chiral symmetry breaking. From the study
of eigenvalue correlations in strongly interacting systems, we have
concluded that they are described naturally with by Random Matrix 
Theory with the global symmetries of the physical system. 
In QCD, this
led to the introduction of chiral Random Matrix Theories. 
They provided us with an analytical understanding of the statistical
properties of the eigenvalues on the scale of a typical level spacing.
In particular,
impressive agreement between lattice QCD and chiral Random Matrix
Theory was found for the microscopic spectral density and for 
spectral correlations in the bulk of the spectrum.
An extension of this model to nonzero chemical potential explains 
some intriguing properties of 
previously obtained lattice QCD Dirac spectra and instanton liquid Dirac 
spectra. 
\subsection{Acknowledgements}

This work was partially supported by the US DOE grant
DE-FG-88ER40388. Poul Damgaard and Jerzy Jurkiewicz are thanked for organizing
this workshop. NATO is acknowledged for financial support. 
We benefitted from discussions with T. Wettig, and 
M. Halasz is thanked for a critical reading of the manuscript. 
Finally, I thank my collaborators on whose work this review is based.

\end{document}